\documentclass[twocolumn,english,prl,floatfix,citeautoscript,nofootinbib]{revtex4}
\usepackage{amsbsy}
\usepackage{latexsym,epsfig,graphicx}
\usepackage{dcolumn}
\usepackage{subfigure}
\usepackage{comment}
\usepackage{color}
\usepackage[colorlinks,urlcolor=blue,citecolor=blue]{hyperref}
\usepackage{amstext}
\usepackage{amssymb}
\usepackage{setspace}

\begin{document}

\title{Spin-twisted Optical Lattices: Tunable Flat Bands and
Larkin-Ovchinnikov Superfluids}
\author{Xi-Wang Luo}
\thanks{xiwang.luo@utdallas.edu}
\author{Chuanwei Zhang}
\thanks{chuanwei.zhang@utdallas.edu}
\affiliation{Department of Physics, The University of Texas at Dallas, Richardson, Texas
75080-3021, USA}

\begin{abstract}
Moir\'{e} superlattices in twisted bilayer graphene and transition-metal
dichalcogenides have emerged as a powerful tool for engineering novel band
structures and quantum phases of two-dimensional quantum materials. Here we
investigate Moir\'{e} physics emerging from twisting two independent
hexagonal optical lattices of atomic (pseudo-)spin states (instead of
bilayers), which exhibits remarkably different physics from twisted bilayer
graphene. We employ a momentum-space tight-binding calculation that includes
all range real-space tunnelings, and show that all twist angles $\theta
\lesssim 6^{\circ }$ can become magic that support gapped flat bands. Due to
greatly enhanced density of states near the flat bands, the system can be
driven to superfluid by weak attractive interaction. Strikingly, the
superfluid phase corresponds to a Larkin-Ovchinnikov state with finite
momentum pairing, resulting from the interplay between flat bands and
inter-spin interactions in the unique single-layer spin-twisted lattice. Our
work may pave the way for exploring novel quantum phases and twistronics in
cold atomic systems.
\end{abstract}

\maketitle

\emph{\textcolor{blue}{Introduction}.---}Twisting two weakly-coupled
adjacent crystal layers has been employed as a powerful tool for tailoring
electronic properties of two-dimensional quantum materials~\cite%
{PhysRevLett.99.256802,nanoLett.10.804,PhysRevB.81.165105,PNAS.108.12233,
PhysRevB.86.155449,PhysRevLett.117.116804,PhysRevLett.119.247402}, such as
the formation of Moir\'{e} superlattices and flat bands. This has been
evidenced by the recent groundbreaking discovery of superconductivity and
correlated insulator phases in twisted bilayer graphene (TBG)~\cite%
{Nature556.43,Nature556.80}, which provide a rich platform for exploring
strongly-correlated many-body phases \cite%
{Nature572.95,science.aaw3780,sciadv.aaw9770,Nat.Phys.15.1174,
Nat.Phys.15.237,PhysRevLett.123.197702}, with the underlying physical
mechanisms still under investigation \cite%
{PhysRevB.98.220504,PhysRevLett.122.257002,PhysRevLett.121.257001,PhysRevLett.121.087001, PhysRevLett.122.026801,PhysRevLett.123.237002,PhysRevB.98.241407,PhysRevB.98.075154,PhysRevLett.121.217001, PhysRevX.8.031089,PhysRevX.8.041041}%
. In TBG, the interactions, the inter- and intra-layer couplings are
generally fixed with very limited tunability~\cite%
{PNAS.114.3364,PhysRevB.98.085144,Nature557.404,science.aav1910}, and magic
flat bands occur only in a narrow range of very small twist angles around $%
\sim 1.1^{\circ }$. Going beyond layer degree of freedom in TBG, two
questions naturally arise. Can lattices of other pseudo degrees be twisted
to realize novel Moir\'{e} lattices and flat bands with great tunability? If
so, can new physics emerge in such twisted systems?

Ultracold atoms in optical lattices provide a promising platform for
exploring many-body physics in clean environments with versatile tunability
\cite{PhysRevLett.81.3108,adv.phys.ultracold,
annurev.Esslinger,nphys2259,RevModPhys.87.637,nature10871,science.aad5812,
nphys1108,3Dclockfermi,clock173Yb,clock87Sr,2D.Materials,arxiv.1809.04604,PhysRevB.101.235121,PhysRevLett.125.030504,RevModPhys.82.1225,alkaline.earth.feshbach}%
. While it is challenging to realize twisted bilayer lattices, the atomic
internal states offer a pseudospin degree, where optical lattice for each
spin state can be controlled independently (in particular for alkaline-earth
atoms) \cite%
{PhysRevA.78.032508,PhysRevLett.101.170504,PhysRevLett.120.143601,PhysRevA.100.053604}%
, allowing the realization of spin-twisted-lattices and related Moir\'{e}
physics. Such spin-twisted-lattices have several remarkable difference from
TBG. For instance, two spins reside on one layer spatially
(instead of bilayer in TBG) with their coupling provided by additional
lasers, resulting in different inter-spin (compared with inter-layer in TBG)
hopping and other physical parameters. The interaction is dominated by the
inter-spin $s$-wave scattering between fermion atoms in relatively twisted spin
lattices, in contrast to the uniform intra-layer interaction
without spin twist in TBG. These differences can significantly affect the
resulting band structures and many-body quantum states. It is unclear
whether extremely flat and gapped bands (i.e., magic-angle behaviors) can
exist in spin-twisted single-layer lattice. If yes, how large can the magic
angle be tuned to? Can new phases emerge from twisted inter-spin
interactions?

In this Letter, we address these important questions by investigating the
Moir\'{e} physics for cold atoms in two spin-dependent hexagonal lattices
twisted by a relative angle, with two spin states coupled by additional
uniform lasers. Our main results are:

\textit{i}) We employ a momentum-space tight-binding method to include all
range real-space tunnelings with high accuracy, which is crucial for
obtaining the correct flat band structures and low-energy physics.

\textit{ii}) Because of the tunability of inter-spin coupling strength and
lattice depth, all twist angles with $\theta \lesssim 6^{\circ }$ can become
magic and support extremely flat and gapped bands. In general, a smaller
magic angle requires weaker inter-spin coupling or a shallower lattice. When
$\theta $ is too large, no flat bands exist in the whole parameter space due
to strong inter-valley coupling.

\textit{iii}) The system can be driven to the superfluid phase by very weak
attractive interactions at magic angles where the flat bands greatly enhance
the density of states (DOS). Strikingly, the superfluid phase corresponds to
a Larkin-Ovchinnikov (LO) state~\cite{LO} with nonzero pairing momentum and
staggered real-space pairing order at the hexagonal lattice scale, which
does not exist in TBG and results from the interplay between flat bands and
the unique inter-spin interactions of atoms in relatively twisted spin
lattices.

\begin{figure}[tb]
\includegraphics[width=1.0\linewidth]{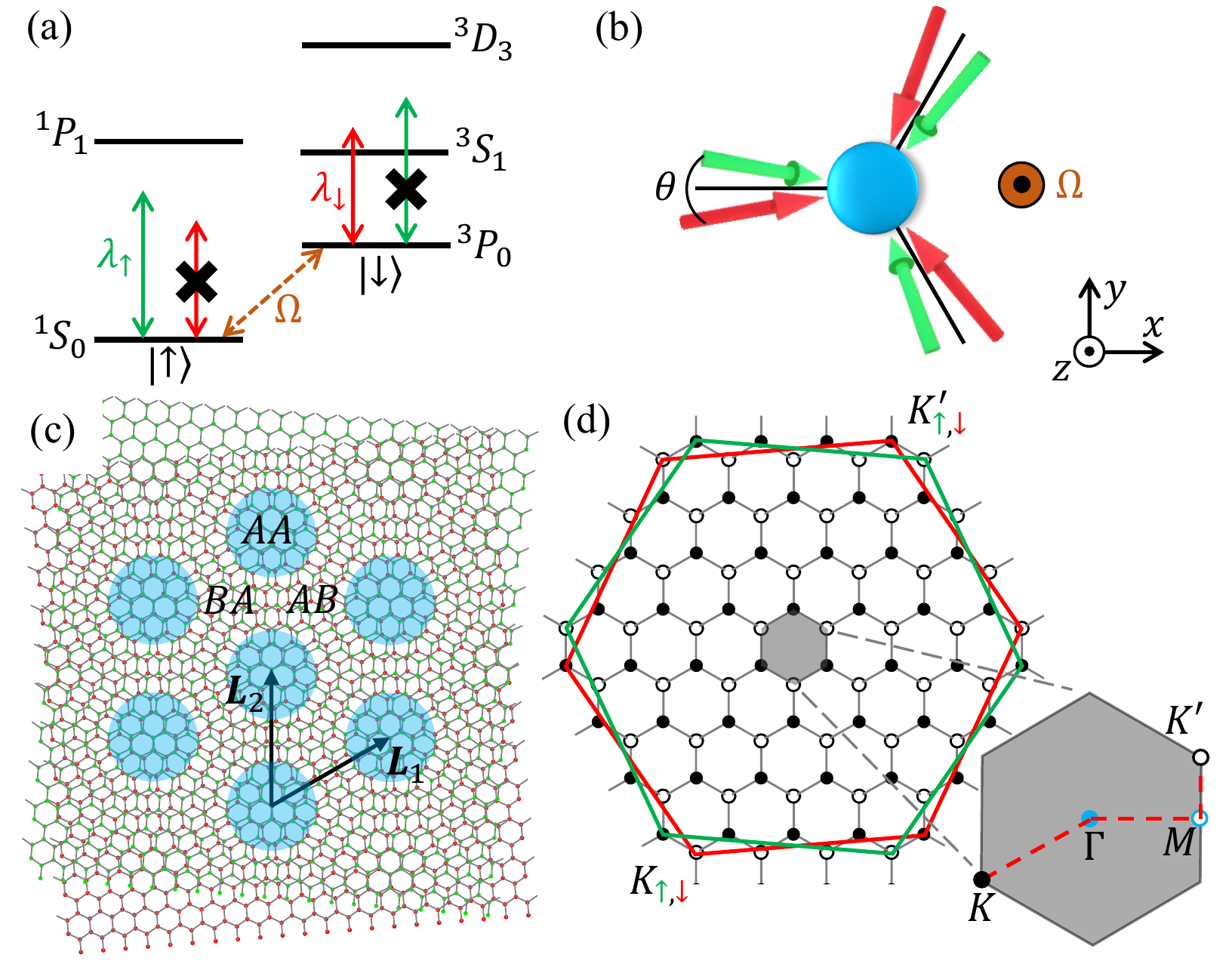}
\caption{(a) Energy level diagram of alkaline-earth(-like) atoms, showing
how state-dependent optical lattices can be realized. (b) Laser
configuration to generate spin-twisted hexagonal lattices. (c) Moir\'{e}
pattern and (d) Brillouin zone of spin twisted hexagonal lattices with $%
\protect\theta =9.43^{\circ }$ ($m=3,n=4$). $AA$ spots form a triangle
lattice with $AB$ or $BA$ spots at the triangles' centers. $\mathbf{L}_{i}$
are the primitive lattice vectors. The large hexagons in (d) correspond to
the bare BZs for states $\uparrow $ (green) and $\downarrow $ (red),
respectively. }
\label{fig:sys}
\end{figure}

\emph{\textcolor{blue}{Model}.---}To obtain independent optical lattices
that can be twisted, we consider two long-lived $^{1}S_{0}$ and $^{3}P_{0}$
orbital states (denoted as pseudospin states $\left\vert \uparrow
\right\rangle $ and $\left\vert \downarrow \right\rangle $) of
alkaline-earth(-like) atoms \cite%
{PhysRevA.78.032508,PhysRevLett.101.170504,PhysRevLett.120.143601,PhysRevA.100.053604}
as shown in Fig.~\ref{fig:sys}a. Atoms in state $\left\vert \uparrow
(\downarrow )\right\rangle $ are trapped solely by $\lambda _{\uparrow
(\downarrow )}$-wavelength lasers, which are tuned-out for atoms in state $%
\left\vert \downarrow (\uparrow )\right\rangle $ (e.g., $\lambda _{\uparrow
,\downarrow }=627\text{nm},689\text{nm}$ for Sr atoms). A hexagonal lattice $%
V(\mathbf{r})=-V_{0}|\sum_{j=1}^{3}\mathbf{\epsilon }_{j}\exp [i\mathbf{k}%
_{L,j}\cdot (\mathbf{r}-\mathbf{r_{0}})]|^{2}$ is generated by intersecting
three lasers at $120^{\circ }$ in the $x$-$y$ plane with each beam linearly
in-plane polarized~\cite{science.aad5812}. Here $V_{0}$ is the trap depth, $%
\mathbf{r_{0}}$ is the hexagonal plaquette center, $\mathbf{k}_{L,j}$ and $%
\mathbf{\epsilon }_{j}$ are the laser wave vector and polarization.
Hereafter, we set momentum and energy units as $k_{R}={2\pi }/{\lambda
_{\downarrow }}$ and $E_{R}={\hbar ^{2}k_{R}^{2}}/{2m}$.
%$\mathbf{k}_{j}=R_z(\frac{2(j-1)\pi}{3})\mathbf{k}_{1}$
%as shown in Fig.~1b.
The two spin-dependent potentials $V_{\uparrow ,\downarrow }(\mathbf{r})$
are obtained through twisting $V(\mathbf{r})$ by $\pm \theta /2$
(see Fig.~\ref{fig:sys}b). The shorter-wavelength $\lambda _{\uparrow }$ lasers
%that are twisted by %an angle
%$\theta$ regarding the $\lambda_{\downarrow}$ lasers,
have an out-of-plane angle to ensure the same lattice constant for two
potentials.
%$V_{\uparrow,\downarrow}(\mathbf{r})=V(\mathbf{r}_{\uparrow,\downarrow})$, with %$\mathbf{r}_{\uparrow,\downarrow}=R_z(\pm\frac{\theta}{2})\mathbf{r}$.
%%as shown in Fig.1.%
%%\begin{equation}
%$V(\mathbf{r})=-V_{0}|\sum^3_{j=1} \mathbf{\epsilon}_j\exp [i\mathbf{k}_{j}\cdot (\mathbf{r}-\mathbf{r_0})]|^{2}$.
%\end{equation}
%with $V_{0}$ the trap depth, $\mathbf{r_0}$ the hexagonal plaquette center,
%$\mathbf{k}_{1}=[{2\pi}/{\lambda_\downarrow},0,0]$ and $\mathbf{\epsilon}_1=[0,1,0]$.
%$\mathbf{k}_{j}=R_z(\frac{2(j-1)\pi}{3})\mathbf{k}_{1}$
%%and $\mathbf{\epsilon}_j=R_z(\theta_j)\mathbf{\epsilon}_1$
%are the wave vectors %and polarizations,
%with $\mathbf{k}_{1}=[{2\pi}/{\lambda_\downarrow},0,0]$ and $R_z(\varphi)$ the rotation about $z$-axis
%by an angle $\varphi$,
%similarly for the polarizations $\mathbf{\epsilon}_j$ with
%$\mathbf{\epsilon}_1=[0,1,0]$.
The $z$-direction is tightly confined by an additional state-independent
potential using the so-called magic-wavelength lasers~\cite%
{RevModPhys.87.637}, which reduces the dynamics to two dimensions (2D).
%The hexagonal
%potentials $V_{\uparrow ,\downarrow }(\mathbf{r})$ have two minima per unit
%cell (corresponding to $A$ and $B$ sublattice sites).
%We start from $AA$
%stacking and twist the two potentials around one of the $A$ sublattice
%sites, as shown in Fig.~\ref{fig:sys}c.
The two pseudospin states
%$\left\vert \uparrow \right\rangle $ and $\left\vert\downarrow \right\rangle $
are coupled by a clock laser~\cite%
{RevModPhys.87.637} propagating along $z$, with $\Omega $ the Rabi
frequency.

%We first consider the
%commensurate twist angles with
%$\cos (\theta )=\frac{n^{2}+m^{2}+4mn}{2(n^{2}+m^{2}+mn)}$
%parameterized by two integers $(m,n)$.
%The Moir\'e
%primitive lattice vectors %are defined as
%$\mathbf{L}_i$ have length
%$|\mathbf{L}_i|=a_0\sqrt{n^{2}+m^{2}+mn}$ with
%$a_0=\frac{2}{3}\lambda_\downarrow$ the hexagonal lattice constant.
We first consider commensurate twists with $\cos (\theta )=\frac{%
n^{2}+m^{2}+4mn}{2(n^{2}+m^{2}+mn)}$ parameterized by two integers $(m,n)$~%
\cite{PhysRevLett.99.256802}. In Figs.~\ref{fig:sys}c and %
\ref{fig:sys}d, the real-space pattern and Moir\'{e} Brillouin zone (BZ) are
shown together with the bare BZs of two spins.
%For typical optical lattice depth, the tunnelings are long-ranged and
%all tunnelings (especially the inter-spin couplings) are
%highly anisotropic (especially for the inter-spin couplings), which
For typical lattice depth, long-range tunnelings beyond nearest neighbors
(especially for the inter-spin couplings where the site separations take
various values and are nearly continuously distributed for small twists)
should be taken into account to obtain the correct magic flat bands~\cite{SM}%
. Small deviations in the tunneling coefficients may result in significant
change in the flat band structures due to the narrow bandwidth. Here we adopt the momentum-space Bloch basis $\{\phi
_{sl\mathbf{k}_{s}}(\mathbf{r})\}$ (with $\mathbf{k}_{s}$ the Bloch
momentum, $l$ the band index and $s=\uparrow ,\downarrow $) of $V_{s}(%
\mathbf{r})$ which spans the same tight-binding Hilbert space as the Wannier
basis. When the two spins are decoupled,
%the Hamiltonian of each spin state reads $%
%H_{s}=\sum_{l,\mathbf{k}_{s}}\mathcal{E}_{sl\mathbf{k}_{s}}\alpha _{sl%
%\mathbf{k}_{s}}^{\dag }\alpha _{sl\mathbf{k}_{s}}$, where $\alpha _{sl%
%\mathbf{k}_{s}}^{\dag }$ is the creation operator corresponding to the Bloch
%state $\phi _{sl\mathbf{k}_{s}}(\mathbf{r})$ of $V_{s}(\mathbf{r})$ with $l$
%the band index and $s=\uparrow ,\downarrow $.
the lowest two bands of each spin state form two Dirac points for $\mathbf{k}%
_{s}$ at valley $K_{s}$ and $K_{s}^{\prime }$ in the bare BZs~\cite{SM}.
%When the two spin states are weakly coupled,

%The two spin states are coupled by a clock laser~\cite%
%{RevModPhys.87.637} propagating along $z$ direction, with $\Omega$ the Rabi frequency.
By projecting onto the %spin-dependent
%tight-binding
basis $\{\phi _{sl\mathbf{k}_{s}}(\mathbf{r})\}$, the inter-spin coupling
Hamiltonian %in momentum space
reads~\cite{SM} %we can project the
%coupling to the tight-binding basis in momentum space,
\begin{equation}
H_{\uparrow \downarrow }(\mathbf{q})=\sum_{l,l^{\prime },\mathbf{g}%
_{\uparrow ,\downarrow }}J_{\mathbf{g}_{\uparrow }\mathbf{g}_{\downarrow
}}^{ll^{\prime }}(\mathbf{q})\alpha _{\uparrow l\mathbf{q}+\mathbf{g}%
_{\uparrow }}^{\dag }\alpha _{\downarrow l^{\prime }\mathbf{q}+\mathbf{g}%
_{\downarrow }}+h.c.,
\end{equation}%
where $\alpha _{sl\mathbf{k}_{s}}^{\dag }$ are the creation operators of the
Bloch states, $\mathbf{q}$ is the superlattice Bloch momentum in the Moir\'{e%
} BZ and $\mathbf{g}_{s}$ are the reciprocal lattice vectors of the Moir\'{e}
superlattice whose summation runs over the bare BZ of state $s$. The
inter-spin coupling coefficients are $J_{\mathbf{g}_{\uparrow }\mathbf{g}%
_{\downarrow }}^{ll^{\prime }}=\langle \phi _{\uparrow l\mathbf{q}+\mathbf{g}%
_{\uparrow }}|\Omega |\phi _{\downarrow l^{\prime }\mathbf{q}+\mathbf{g}%
_{\downarrow }}\rangle $, which already incorporate all range real-space
tunnelings. Another advantage of this momentum-space approach is that if
only the low-energy physics is of interest, then we only need to keep $l$
and $\mathbf{g}_{s}$ that correspond to the low-energy Bloch states~\cite%
{PhysRevLett.99.256802,PhysRevB.81.165105,PNAS.108.12233,PhysRevB.86.155449}%
, leading to a rather rapid convergence of the basis set.

Although spin-twisted optical lattices share some similarities with TBG,
several important differences need be noted: 1) The two twisted optical
potentials are spin-dependent and do not affect each other, while in TBG
electrons in one layer can feel the potential of the other layer; 2) The
inter-spin couplings in the single layer (realized by additional lasers) are
different from the inter-layer tunnelings in TBG~\cite%
{PhysRevLett.99.256802,SM}; 3) The optical lattice potential takes a simple
cosine form, therefore the bare bands and inter-spin couplings can be
obtained accurately from the Bloch states, while TBG Hamiltonians are
usually based on real-space tight-binding approximation expressed in
Slater-Koster parameters~\cite%
{PhysRevLett.99.256802,PhysRev.94.1498,JPSJ.70.1647,PhysRevB.85.195458,PhysRevB.93.235153}%
; 4) Long-range tunnelings are more significant due to shallow lattices
considered here, which not only improve the atomic lifetime, but also
increase the bare Dirac velocity. 5) Interactions are dominated by the $s$%
-wave scattering between fermion atoms in relatively twisted lattices, while
electronic interactions in TBG, including both Coulomb repulsive
and phonon-mediated attractive interactions, mainly involve electrons in the
same layer with no relative twist~\cite%
{PhysRevB.98.220504,PhysRevLett.122.257002,PhysRevLett.121.257001,
PhysRevX.8.041041,PhysRevLett.121.217001,PhysRevLett.122.026801}; 6)
Finally, the cold-atom parameters (e.g., inter-spin tunnelings, lattice
depth, lattice constant, interactions, etc.) are highly tunable, comparing
with one tunable parameter, twist angle, in TBG.

\begin{figure}[tb]
\includegraphics[width=1.0\linewidth]{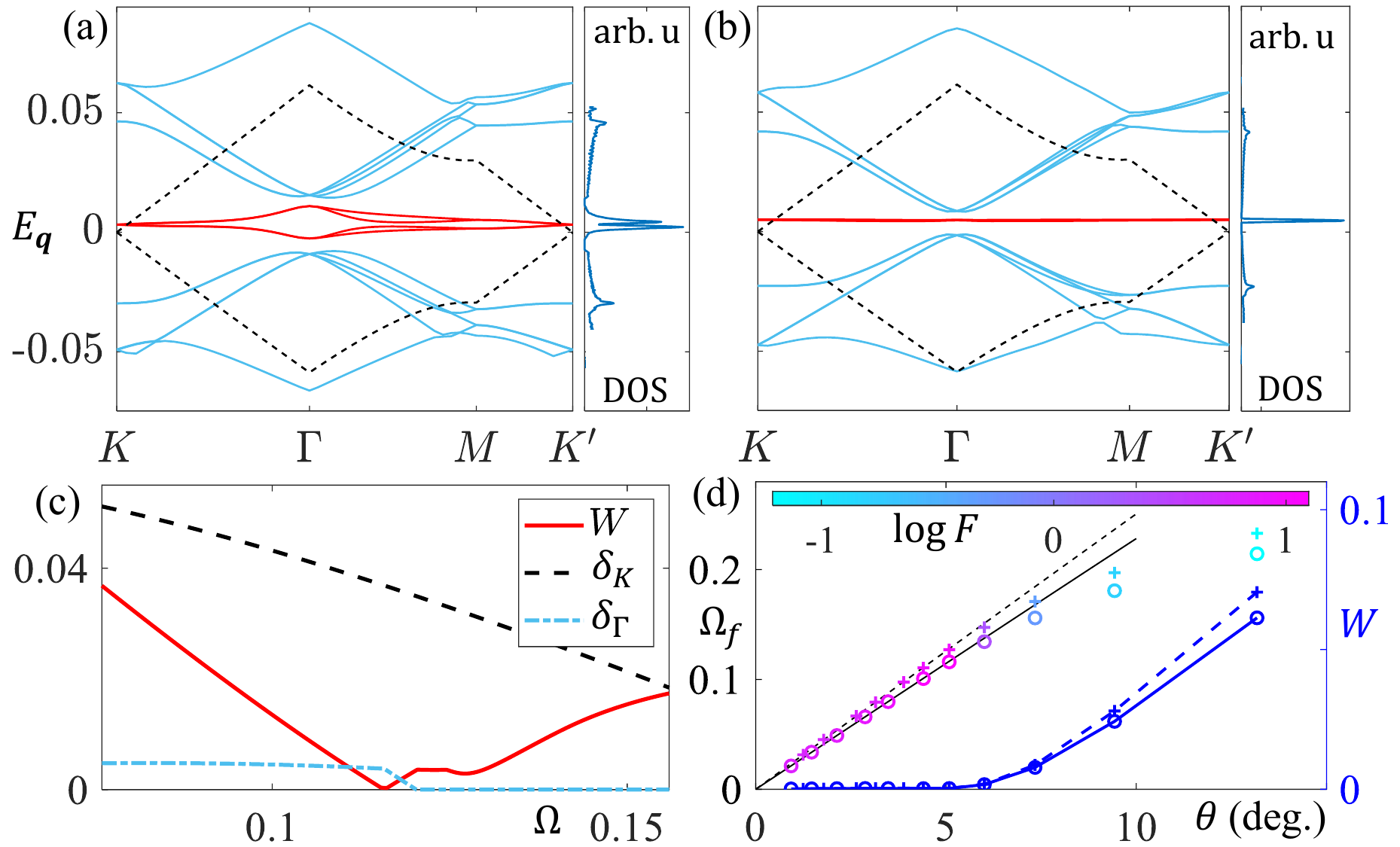}
\caption{(a) and (b) Moir\'{e} bands along high-symmetry lines (the red
dashed lines in Fig.~\protect\ref{fig:sys}d) and DOS for $\Omega =0.1$ and $%
\Omega =0.116$, respectively. We set the bare Dirac cone energy
as zero. The black dashed lines are bare Dirac bands folded back to Moir\'{e%
} BZ. (c) Flat band width $W$ and gaps $\protect\delta _{K,\Gamma }$ with
other higher bands at $K$ and $\Gamma $ points. In (a)-(c), $\protect\theta %
=5.086^{\circ }$ and $V_{0}=6$. (d) Critical coupling $\Omega _{f}$ as a
function of $\protect\theta $ with $V_{0}=6$ (circles) and $V_{0}=4$ (plus
signs). Color bars show the flatness at $\Omega =\Omega _{f}$ with flat band
width shown by the thick blue markers and lines. The thin solid (dashed)
line corresponds to $c =1.932$ at $V_{0}=6$ ($c %
=1.827$ at $V_{0}=4$). }
\label{fig:bands}
\end{figure}

\emph{\textcolor{blue}{Flat bands}.---}We solve the Moir\'{e} bands
numerically and find that all small twist angles ($\theta \lesssim 6^{\circ }
$) can become magic that support flat bands with proper choice of inter-spin
coupling strength or lattice depth. %The Moir\'e
%primitive lattice vectors %are defined as
%$\mathbf{L}_i$ have length
%$|\mathbf{L}_i|=a_0\sqrt{n^{2}+m^{2}+mn}$ with
%$a_0=\frac{2}{3}\lambda_\downarrow$ the hexagonal lattice constant.
In Figs.~\ref{fig:bands}a and \ref{fig:bands}b, we plot the band structures
for different inter-spin coupling strengths $\Omega $ with $V_{0}=6$ and $%
\theta =5.086^{\circ }$ ($m=6,n=7$). Similar to the TBG, the system has four
low-energy bands, two of which form a Dirac cone at the Moir\'{e} $K$ ($%
K^{\prime }$) point where the remaining two bands are split by a tiny gap
due to the inter-valley ($K_{s}$-$K_{\bar{s}}^{\prime }$) coupling. The
Dirac cones shift to a higher energy compared to the bare ones, which is due
to the couplings with states away from the valleys that have weak
nonlinearity in the dispersion.
%the weak nonlinearity in dispersion away from the valleys;
%Such effects are more significant for larger twist angles and shallower
%lattices. % (with stronger long-range tunnellings).
%the long-range $AA$ or $BB$ tunnellings strongly break the sublattice (chiral) symmetry of each spin state and
%the two bare Dirac bands are asymmetric with respect to Dirac-point energy.
%Nevertheless,
The inter-spin coupling reduces the Dirac velocity significantly and enhance
the DOS near the Dirac cones, as shown in Fig.~\ref{fig:bands}a. The peaks
in the DOS correspond to the Van Hove singularities near the Moir\'{e} $M$
points~\cite{PhysRevLett.122.026801,Nat.Commun.10.5769}.
%Continuously increasing $\Omega$ would further reduce
The bandwidth $W$ of the low-energy bands and Dirac velocity are reduced
further as $\Omega $ increases and may even vanish (i.e., the twist angle
becomes magic) at certain inter-spin coupling strength. We are interested in
the flat bands associated with magic angles and will focus on the physics
around the critical coupling $\Omega _{f}$ where the narrowest bandwidth
occurs (as shown in Fig.~\ref{fig:bands}b). For $\Omega \lesssim \Omega _{f}$%
, the four low-energy bands are always separated by an energy gap from other
bands in the spectrum, the gap is minimized near the Moir\'{e} $\Gamma $
point and would close eventually as we increase $\Omega $ above $\Omega _{f}$%
. Shown in Fig.~\ref{fig:bands}c are the bandwidth $W$ and gaps $\delta
_{\Gamma ,K}$ (with other higher bands) versus %as functions of
$\Omega $. % would close the gap.
%the Dirac velocity would be reduced and
%may even vanish at the `magic angle' (for a given inter-spin coupling strength).

In Fig.~\ref{fig:bands}d, we plot $\Omega _{f}$ and the corresponding
bandwidth $W$ and flatness $F\equiv \delta _{\Gamma }/W$ as functions of the
twist angle $\theta $. %where $F$
%and $\delta _{\Gamma }$ is the
%gap of the flatbands with other higher bands at Moir\'{e} $\Gamma $
%point.
For small twists, the low-energy bands are mainly determined by the
states with $\mathbf{g}_{s}$ around the Dirac valleys, and have a
narrow width and high flatness at $\Omega =\Omega _{f}$. In addition, the
inter-valley coupling is weak, %at small $\theta$ %twist angles
thus two conduction or valence bands (one from each valley) are nearly
degenerate along the high-symmetric $\Gamma $-$K$ ($K^{\prime }$) lines~\cite%
{PhysRevLett.122.026801}. %Also,
We find $\Omega _{f}$ almost linearly increases with $\theta $.
%and the dependence is
%almost linear for
%$\theta\lesssim6^\circ$. %
%small twist angles.
Specifically, the magic flat bands occur near $c =$ const., where $%
c \equiv \frac{\Omega }{v_{D}k_{D}}$ is a dimensionless parameter with $%
k_{D}=2k_{R}\sin (\theta /2)$ the $K$-$K^{\prime }$ distance in Moir\'{e} BZ
and $v_{D}$ the bare Dirac velocity. % $v_D=0.79 V0=4;0.68 V0=6$
%Defining a dimensionless parameter
%$\alpha\equiv\frac{\Omega}{v_Dk_D}$
%[$k_D=k_R\sin(\theta/2)$ is $K$-$K'$ distance in MBZ],
%the magic flatbands
%occur near constant $\alpha=1.15$,
%(i.e., $\frac{\Omega_f}{v_Dk_D}\simeq1.15$),
%with $\alpha\equiv\frac{\Omega}{v_Dk_D}$
%a dimensionless parameter
%$\alpha=\frac{\Omega}{v_Dk_D}$ with
%and ,
This is consistent with the continuum model in the TBG where $c $ is
the single parameter~\cite{PNAS.108.12233,PhysRevB.86.155449}. When the
twist angles are large $\theta >6^{\circ }$, the width and splitting of the
four low-energy bands become comparable or larger than the gap with other
bands, and no magic flat bands exist for any $\Omega $ since the
inter-valley couplings and the effects of states %with $\mathbf{g}_s$
away from the bare Dirac valleys become significant. %, therefore,
%the four low-energy bands have large splitting
%and no magic flatbands exist for any
%$\Omega$.
% and $V_0$.
%at small twists.
%For a small twist angle,
%,  for small $\theta$.
For incommensurate twist angles, we can generalize the continuum model and
only keep $\mathbf{g}_{s}$ around one valley, which should be valid for
small $\theta $~\cite{SM}.
%where the inter-valley coupling and states far away from the
%valley are negligible in determining the low-energy bands~\cite{SM}.
We thus conclude that all small angles $\theta \lesssim 6^{\circ }$ can
support magic flat bands. %, as shown in Fig.~2d.

For different lattice depths $V_{0}$, the magic behaviors discussed above
are similar (see Fig.~\ref{fig:bands}d). Meanwhile, a smaller $V_{0}$ leads
to a larger $v_{D}$ and thereby a stronger $\Omega _{f}$ (for fixed $\theta $%
). Long-range tunnelings are also more significant in a shallower lattice,
which would effectively enhance the inter-spin couplings, leading to a
slightly smaller $c$ where the flat bands occur. The flatness may also
be improved by decreasing $V_{0}$ properly, since a larger $v_{D}$ leads to
a larger gap $\delta _{\Gamma }$~\cite{PNAS.108.12233,PhysRevB.86.155449}
and long-range tunnelings in real space can reduce inter-valley couplings
that have large momentum separations. However, in the very shallow region
where the dispersion-linearity around the bare Dirac cone becomes poor, the
flatness starts to decrease with $V_{0}$.
%if the lattice becomes too shallow where
%the sublattice symmetry is strongly broken
%for each spin state and
%the nonlinearity
%in dispersion around the bare Dirac cone is notable,
%the flatness becomes poor.

\begin{figure}[tb]
\includegraphics[width=1.0\linewidth]{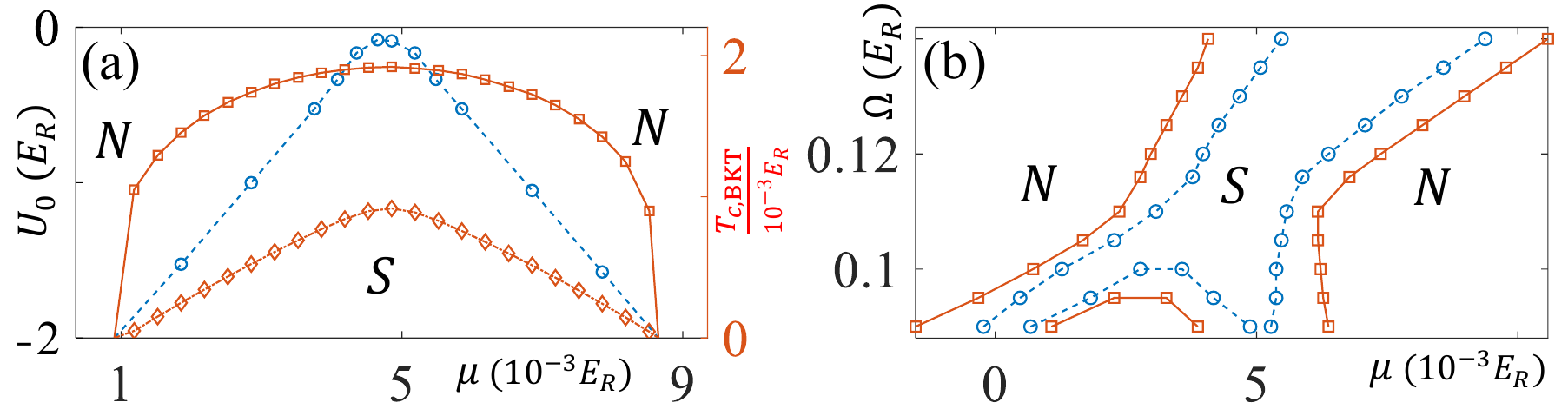}
\caption{(a) Phase diagrams in the $U_{0}$-$\protect\mu $ plane at zero
temperature (blue dots). The critical temperatures $T_{c}$ (red squares) and
$T_{\text{BKT}}$ (red diamonds) as functions of $\protect\mu $ at $U_{0}=-2$%
, with $\Omega =\Omega _{f}$. (b) Zero-temperature phase diagrams in the $%
\Omega $-$\protect\mu $ plane for $U_{0}=-0.5$ (blue dots) and $U_{0}=-1$ (red
squares). $N$ and $S$ represent the normal and superfluid phases,
respectively.
%We have set the bare Dirac cone energy as zero point ($\protect\mu =0$).
Common parameters: $\protect\theta =5.086^{\circ }$, $V_{0}=6$.}
\label{fig:phase}
\end{figure}

\begin{figure}[tb]
\includegraphics[width=1.0\linewidth]{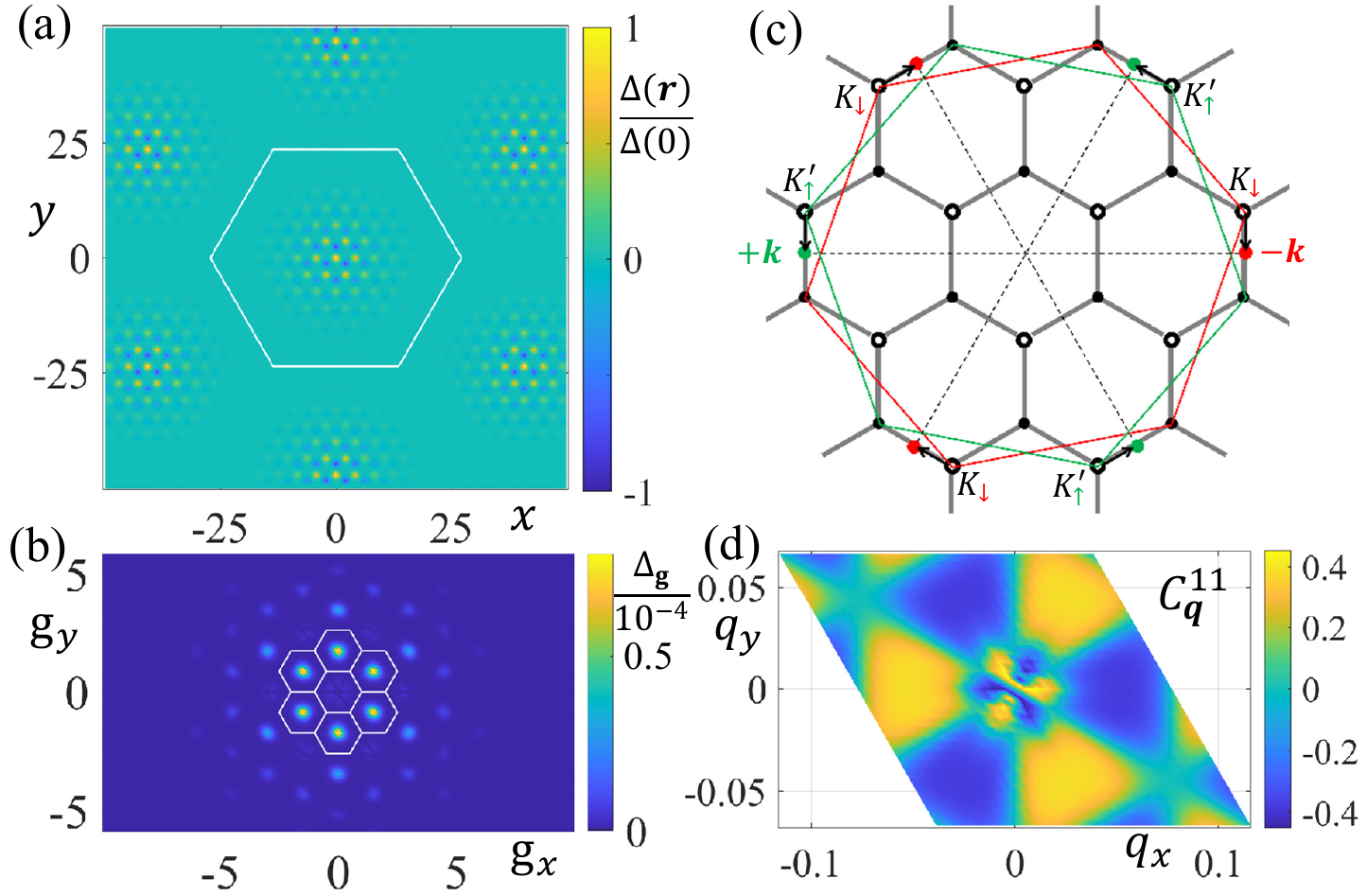}
\caption{(a) and (b) The superfluid pairing amplitudes in real [$\Delta (%
\mathbf{r})$] and momentum space ($\Delta _{\mathbf{g}}$), respectively. The
white hexagons correspond to the Moir\'{e} unit cell in (a) and the
untwisted bare BZs in (b). (c) Schematic illustration of the pairing
(indicated by dashed lines) in the BZs. (d) The correlation $C_{%
\mathbf{q}}^{11}$. Common parameters: $\protect\theta =5.086^{\circ }$, $%
V_{0}=6$, $\Omega =\Omega _{f}$ and $U_{0}=-0.5$.}
\label{fig:order}
\end{figure}

\emph{\textcolor{blue}{Superfluid orders}.---}The narrowly dispersing flat
bands suppress the kinetic energy and atom-atom interactions can lead to
strongly correlated many-body ground states. Different from TBG~\cite%
{PhysRevB.98.220504,PhysRevLett.122.257002,PhysRevLett.121.257001,
PhysRevX.8.041041,PhysRevLett.121.217001,PhysRevLett.122.026801}, here the
interaction of fermion atoms is dominated by $s$-wave scattering between atoms in relatively
twisted lattices, with strength tunable through Feshbach resonance~\cite%
{RevModPhys.82.1225,alkaline.earth.feshbach},
%We focus on the superfluid orders
%driven by attractive $s$-wave scattering % the interaction reads
\begin{equation}
H_{\text{int}}=U_{0}\int d^{2}\mathbf{r}\hat{\Psi}_{\uparrow }^{\dag }(%
\mathbf{r})\hat{\Psi}_{\downarrow }^{\dag }(\mathbf{r})\hat{\Psi}%
_{\downarrow }(\mathbf{r})\hat{\Psi}_{\uparrow }(\mathbf{r}).
\end{equation}%
We are interested in the superfluid order driven by attractive interactions.
We adopt the mean-field approach~\cite%
{PhysRevB.98.220504,PhysRevLett.122.257002,PhysRevLett.121.257001} with
local pairing amplitude $\Delta (\mathbf{r})=U_{0}\langle \hat{\Psi}%
_{\downarrow }(\mathbf{r})\hat{\Psi}_{\uparrow }(\mathbf{r})\rangle $, and
assume that it has Moir\'{e} periodicity~\cite{PhysRevLett.121.257001},
which can therefore be expanded in the form $\Delta (\mathbf{r})=\sum_{%
\mathbf{g}}\Delta _{\mathbf{g}}e^{i\mathbf{g}\cdot \mathbf{r}}$ with $%
\mathbf{g}$ the Moir\'{e} reciprocal lattice vectors.
%The gap equation is~\cite{SM}
%\begin{equation}
%\Delta _{\mathbf{g}}=U_{0}\sum_{j^{\prime },j,\mathbf{q}}\chi _{j^{\prime
%}j}^{\mathbf{q}}(\mathbf{g})C_{\mathbf{q}}^{j^{\prime }j}[\Delta (\mathbf{r}%
%)],  \label{eq:Gap}
%\end{equation}%
%where $j$, $j^{\prime }$ are the Moir\'{e} band labels. The correlation $C_{%
%\mathbf{q}}^{j^{\prime }j}=\langle \beta _{j^{\prime }-\mathbf{q}}\beta _{j%
%\mathbf{q}}\rangle $ can be obtained from the BdG equation and $\chi
%_{j^{\prime }j}^{\mathbf{q}}(\mathbf{g})=\frac{1}{L}\int d^{2}\mathbf{r}e^{-i%
%\mathbf{g}\cdot \mathbf{r}}\psi _{j^{\prime }-\mathbf{q}}(\mathbf{r}%
%,\downarrow )\psi _{j\mathbf{q}}(\mathbf{r},\uparrow )$ with $L$ the system
%volume, $\psi _{j\mathbf{q}}$ the Moir\'{e} wavefunction and $\beta _{j%
%\mathbf{q}}$ the corresponding annihilation operator.
We use the Bogoliubov-de Gennes Hamiltonian to obtain $\Delta (\mathbf{r})$
self-consistently~\cite{SM}, and retain only the four flat bands that have
much larger DOS than nearby bands. We have verified that the physics is
hardly affected by numerically including more nearby bands~\cite{SM}.

The phase diagrams for $\theta =5.086^{\circ }$, $V_{0}=6$ and $\Omega
=\Omega _{f}$ are shown in Fig.~\ref{fig:phase}a. Due to the greatly
enhanced DOS near the magic flat bands at $\Omega _{f}$, the system could be
driven to superfluid by very weak attractive interaction $|U_{0}|\lesssim
0.08$ (at zero temperature) when the chemical potential %$\mu \simeq 0.005$
matches the flat band energy. As $\mu $ is tuned away from flat bands, the
required interaction strength for superfluid phase increases (almost
linearly). For a moderate interaction strength, the mean-field critical
temperature $T_{c}$ could be relatively high (reaches its largest value at $%
\mu \simeq 0.005$) and shows a similar behavior as that predicted in TBG
system~\cite{PhysRevLett.121.257001}.

Note that at finite temperature, the relevant physics in 2D is the
Berezinskii-Kosterlitz-Thouless (BKT) transition~\cite%
{berezinskii1971destruction,kosterlitz1972long,PhysRevLett.109.105302,PhysRevLett.114.110401}
because no long-range superfluid order exists due to phase
fluctuations, and the mean-field $T_{c}$ is often overestimated. The BKT
critical temperature $T_{\text{BKT}}$ could be obtained from the mean-field
superfluid weight~\cite{PhysRevLett.112.086401,julku2018superfluid,SM},
which is numerically calculated with the results shown in Fig.~\ref%
{fig:phase}a (roughly, $T_{\text{BKT}}\simeq 0.4T_{c}$).

In Fig.~\ref{fig:phase}b, we plot the phase diagrams in the $\Omega $-$\mu $
plane. Away from $\Omega _{f}$, the bandwidth will be broadened, and the
superfluid area becomes wider. However, it requires a lower critical
temperature or stronger interaction due to the reduced DOS. At the $\Omega
<\Omega _{f}$ side, the flat band DOS peak splits into two peaks
(corresponding to the Van Hove singularities near the Moir\'{e} $M$ points),
therefore the superfluid phase also splits into two regions where $\mu $
matches the DOS peaks. At the $\Omega >\Omega _{f}$ side, the DOS peak is
simply broadened. As the $|U_{0}|$ decreases, the superfluid phase shrinks
to the area around $\Omega \simeq \Omega _{f}$ and $\mu \simeq 0.005$.

Strikingly, we find that the superfluid phase corresponds to a LO state~\cite%
{LO}, which is very different from that in TBG. The Cooper pairs have
nonzero center-of-mass momentum with $\Delta _{\mathbf{g}}$ mainly
distributed around the first reciprocal lattice vector shell of the
untwisted hexagonal lattice and nearly vanishing around zero momentum,
leading to the staggered real-space pairing orders at the hexagonal lattice
scale (Figs.~\ref{fig:order}a and \ref{fig:order}b).
%as shown in Fig.~\ref{fig:order}b.
The attractive $s$-wave interaction pairs atoms from opposite valleys, and
the superfluid order is peaked in the $AA$ regions, where the local DOS for
the flat bands is strongly concentrated~\cite{SM} and the wavefunction
overlap between two spin states is significant. Therefore, the
intra-sublattice pairing is dominant. Because atoms at the same sublattices
and opposite valleys share opposite angular momenta under the threefold
rotation, the pairing order has the same phase factor for the same
sublattice.

Moreover, the pairing is between Moir\'{e} states at $\pm \mathbf{q}$, which
are mainly determined by the bare Bloch states $\phi _{s l\pm\mathbf{k}}$
%and $\phi _{\downarrow l(-\mathbf{k)}}$
at $\pm \mathbf{k}$ nearest
to the valleys (thereby contributing most to the flatbands). In Fig.~\ref%
{fig:order}c, the pairing between $\uparrow $ states (green dots at $+%
\mathbf{k}$) around valley $K_{\uparrow }^{\prime }$ and $\downarrow $
states (red dots at $-\mathbf{k}$) around valley $K_{\downarrow }$ is
illustrated schematically. Due to the relative twist, $\pm \mathbf{k}$ are
at the same side of $K_{\uparrow }^{\prime }$ and $K_{\downarrow }$,
respectively (see the black arrows in Fig.~\ref{fig:order}c).
Therefore, we have $\phi _{\uparrow l\mathbf{k}}\propto \lbrack
1,e^{i\gamma _{\uparrow \mathbf{k}}}]^{T}$ and $\phi _{\downarrow l\left( -%
\mathbf{k}\right) }\propto \lbrack 1,e^{i\gamma _{\downarrow \left( -\mathbf{%
k}\right) }}]^{T}$ on the $A$ and $B$ sublattice basis, with $\gamma
_{\uparrow \mathbf{k}}\simeq -\gamma _{\downarrow \left( -\mathbf{k}\right)
}+\pi $.
The relative phases $\gamma _{s\mathbf{k}}$
are related to the chirality of the valleys
(i.e., the Berry phase on loops surrounding the valley),
%$\gamma _{s\mathbf{k}}$
%between $A$ and $B$ sublattice sites
which are responsible for the staggered pairing order $\Delta (\mathbf{r})\propto \langle \phi
_{\uparrow l\mathbf{k}}\phi _{\downarrow l\left( -\mathbf{k}\right) }\rangle
\propto \lbrack 1,-1]^{T}$ \cite{SM}. Such LO order is unique for
spin-twisted system with pairing between atoms from relatively twisted
lattices. In TBG, the pairing between spin-up and -down electrons in the
same layer (with no relative twist) leads to ordinary BCS order~\cite%
{PhysRevLett.122.257002,PhysRevLett.121.257001}.

%The relative $\pi$-phase between different
%sublattices of the pairing order results from
%the unique interaction between relatively twisted atoms.
% which is a result of the interplay between the unique
%Though with similar single particle behaviors as in twisted bilayer system,
%the
%inter-pseudolayer interactions and the flat bands in such physically
%single-layer lattices.

%The correlation $C_{\mathbf{q}}^{j^{\prime }j}=\langle \beta _{j^{\prime }-\mathbf{q}}\beta _{j\mathbf{q}}\rangle $

The correlation $C_{\mathbf{q}}^{j^{\prime }j}=\langle \beta _{j^{\prime }-%
\mathbf{q}}\beta _{j\mathbf{q}}\rangle $ shows $f$-wave structure ($\beta _{j\mathbf{q}}$ is the annihilation operator for
the $j$-th flatband),
their combined effects lead to the
nearly uniform superfluid gap~\cite{SM} and the
pairing is $s$-wave.
%The pairing is mainly between bands from different valleys,
The valence bands from different valleys become degenerate along the high
symmetric $\Gamma $-$K$ lines with avoided crossing (a tiny gap) due to
inter-valley couplings, therefore $C_{\mathbf{q}}^{11}$ changes from
characterizing $K_{s}$-$K_{\bar{s}}^{\prime }$ to $K_{\bar{s}}^{\prime }$-$%
K_{s}$ correlations across the $\Gamma $-$K$ lines where its sign flips
(see Fig.~\ref{fig:order}d).
%Though all $C_{\mathbf{q}}^{j^{\prime }j}$
%varies strongly in the Moir\'{e} BZ, their combined effects lead to the
%nearly uniform superfluid gap~\cite{SM}.

%\emph{\textcolor{blue}{Experimental consideration}.---} A larger twist would
%support a larger gap Here long-range tunnelings in real space are very
%significant, since shallow lattices should be used to reduce the atomic loss
%rate and improve the bare Dirac velocity $v_D$ for better isolation of the
%flat bands under twist (as can be seen later).

\emph{\textcolor{blue}{Discussion and conclusion}.---}%
%Due to the high tunability of the cold atom system,
The `magic-angle' physics in the spin-twisted optical lattice is
%different from TBG system in many
%aspects~\cite{SM},
very robust, supporting magic flat bands and novel LO superfluid order in a
wide range of parameter space ($\theta $, $V_{0}$, $\Omega $, $U_{0}$, etc).
%which can be realized in experiments.
For $\theta \simeq 5^{\circ }$ and $V_{0}=6$, the gap between flat bands and
other bands is $\sim 10^{-2}E_{R}$ (about tens of Hz for Sr atoms) and can
be improved further using shallower lattices (larger $v_{D}$) or larger
twists. %Therefore,
The flat bands and enhanced DOS can be observed within atomic gas lifetime
(a few %at the order of %can be up to
seconds for the shallow lattice considered here) using spectroscopic
measurements (e.g., radio-frequency spectroscopy)~\cite%
{science.1100818,nature07172,nature07176,zhang2014fermi}. The critical
superfluid temperature $T_{c,
\text{BKT}}$ is in the nanokelvin region ($\sim
10^{-3}E_{R}$) which might be possible with the recently developing
cold-atom cooling techniques~\cite%
{annurev.Esslinger,science.1236362,nature22362,science.aaz6801}. Thanks to
the large twist angle $\theta \lesssim 6^{\circ }$, the Moir\'{e} unit-cell
may contain less than 100 hexagons; therefore, the magic phenomena can be
observed using a small system with tens of hexagons along each direction.
%For small twist angels with flatbands,
The magic-angle physics is similar for different stackings or twist axes~%
\cite{SM}.

In summary, we study the Moir\'{e} flat band physics and the associated
superfluid order in spin-twisted optical lattices for ultracold atoms, which
showcase magic-angle behaviors for continuum of twists up to $6^{\circ }$
and novel LO superfluid phase remarkably different from that in TBG.
%which showcase remarkably different physics from TBG.
% with nonzero
%center-of-mass momentum Cooper pairs and
%staggered pairing
%amplitudes at the hexagonal lattice
%scale.
%The cold atom system is highly tunable,
In future, it would be interesting to study spin-twisted lattices of other
types (square, triangle, etc) or with different lattice depth and gapped
bands (similar as transition metal dichalcogenide based Moir\'{e} systems~%
\cite{PhysRevLett.121.026402,Nature579.353}). Moreover, one could study
possible interesting many-body states under repulsive interaction and may
even consider the nuclear spin states of alkaline-earth atoms with both
nuclear-spin-exchange and inter-spin interactions.
%In future, it would be interesting to study possible strongly correlated
%states under repulsive interactions, or with gapped bands (i.e., similar as
%transition metal dichalcogenide based Moir\'{e} systems~\cite%
%{PhysRevLett.121.026402,Nature579.353}). Other types of spin-twisted
%lattices (square, triangle, etc.) or twisting two lattices with different
%depths may induce different band structures and novel physics. Moreover, one
%may consider the nuclear spin states of alkaline-earth atoms which should
%lead to interesting many-body physics due to the %highly tunable
%nuclear-spin-exchange and inter-spin interactions. %also include
%the state-dependent trap depth or period,
%as well as
In all, our work provides a highly tunable playground for exploring quantum
many-body physics and twistronics with novel twisted pseudo degrees of
freedom.

%
%Here long-range tunnelings in real space are very
%significant, since shallow lattices should be used to reduce the atomic loss
%rate and improve the bare Dirac velocity $v_D$ for better isolation of the
%flat bands under twist (as can be seen later).

\begin{acknowledgments}
\textbf{Acknowledgements}: We thank J.~H.~Pixley for helpful discussions.
This work is supported by AFOSR (FA9550-16-1-0387, FA9550-20-1-0220), NSF
(PHY-1806227), and ARO (W911NF-17-1-0128).
\end{acknowledgments}

\newpage

\begin{widetext}
\setcounter{figure}{0} \renewcommand{\thefigure}{S\arabic{figure}} %
\setcounter{equation}{0} \renewcommand{\theequation}{S\arabic{equation}}

\section{Supplementary Materials}

\subsection{Momentum-space tight-binding characterization}

The Bloch states of the hexagonal lattice $V(\mathbf{r})=-V_{0}|%
\sum_{j=1}^{3}\mathbf{\epsilon }_{j}\exp [i\mathbf{k}_{L,j}\cdot (\mathbf{r}-%
\mathbf{r_{0}})]|^{2}$ can be written as $\phi _{l\mathbf{k}}(\mathbf{r}%
)=e^{i\mathbf{k}\cdot \mathbf{r}}u_{l\mathbf{k}}(\mathbf{r})$. The periodic
part can be expanded as $u_{l\mathbf{k}}(\mathbf{r})=\sum_{\mathbf{p}}c_{l%
\mathbf{k}}^{\mathbf{p}}e^{i\mathbf{p}\cdot \mathbf{r}}$, where $\mathbf{p}%
=p_{1}\mathbf{e}_{1}+p_{2}\mathbf{e}_{2}$ with $\mathbf{e}_{i}$ the
primitive reciprocal lattice vectors and $p_{i}$ integers. By substituting
expansion of $\phi _{l\mathbf{k}}(\mathbf{r})$ into the Schr\"{o}dinger
equation $[-\frac{\mathbf{\nabla }^{2}}{2m}+V(\mathbf{r})]\phi _{l\mathbf{k}%
}=\mathcal{E}_{l\mathbf{k}}\phi _{l\mathbf{k}}$, the Bloch states (i.e., the
coefficients $c_{l\mathbf{k}}^{\mathbf{p}}$) and bands can be obtained. The
bare Bloch bands $\mathcal{E}_{sl\mathbf{k}_{s}}$ and states $\phi _{sl%
\mathbf{k}_{s}}$ of the two pseudospin states can be obtained similarly, as
shown in Fig.~\ref{fig:S1}. Here we keep the expansion coefficients up to $%
p_{i}=\pm 6$ in the calculation. We see that the two lowest bands (i.e., $%
\mathcal{E}_{sl\mathbf{k}_{s}}$ with $l=1,2$), which have a gap from higher
bands, form two Dirac cones at valleys $K_{s}$ and $K_{s}^{\prime }$~\cite%
{science.aad5812S}. For typical lattice depths, the two Dirac bands are
asymmetric with respect to Dirac-point energy $E=0$ due to the long-range
tunnelings beyond nearest neighbors that break the sublattice symmetry. Such
effect is more significant for shallower lattices.

\begin{figure}[b]
\includegraphics[width=0.85\linewidth]{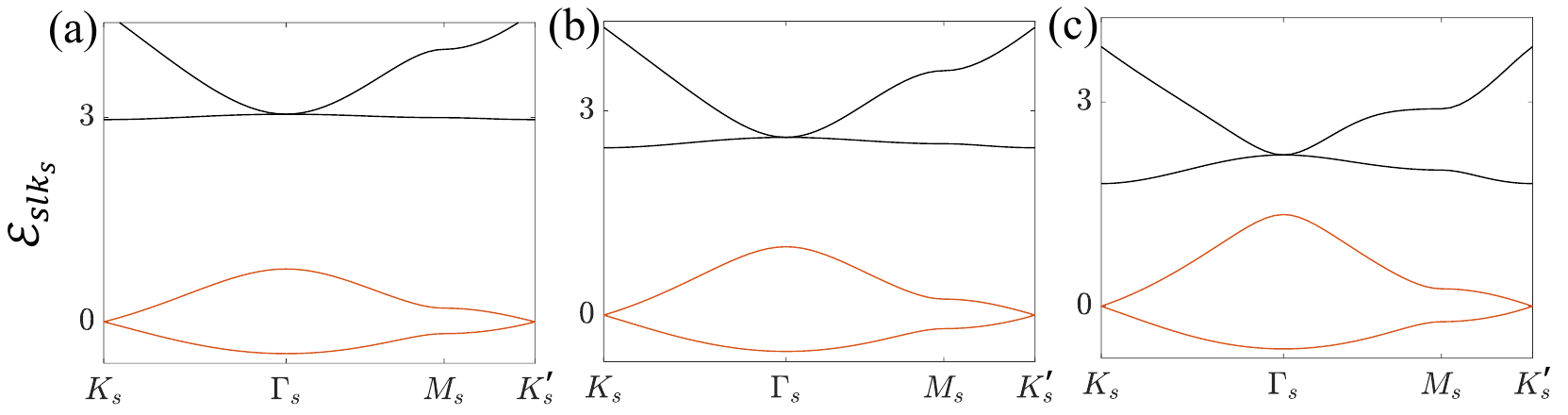}
\caption{Band structures of a hexagonal lattice with (a) $V_{0}=6E_{R}$, (b)
$V_{0}=4E_{R}$ and (c) $V_{0}=2E_{R}$. We have shifted the Dirac-point
energy to $E=0$. For a shallower lattice, the tunnelings (both short and
long range tunnelings) are stronger, therefore, the Dirac velocity is larger
and the nonlinearity of the dispersion around the Dirac points is stronger.
In addition, the two Dirac bands become more asymmetric with respect to the
Dirac-point energy $E=0$ due to the enhanced long-range tunnelings that
break the sublattice symmetry.}
\label{fig:S1}
\end{figure}

We are interested in the low-energy physics near the Dirac points, therefore
we only retain the two-lowest bare Dirac bands (i.e., $\mathcal{E}_{sl%
\mathbf{k}_{s}}$ with $l=1,2$) and drop all higher bands with $l>2$ safely.
That is, we keep only one Wannier orbital at each hexagonal lattice site,
and these Wannier orbitals form the full tight-binding basis set. We find
that it is more convenient to work in the Bloch basis $\{\phi _{sl\mathbf{k}%
_{s}}\}$, which is equivalent to the Wannier basis up to a Fourier
transformation. For commensurate twists $\cos (\theta )=\frac{n^{2}+m^{2}+4mn%
}{2(n^{2}+m^{2}+mn)}$, one can fold the bare Brillouin zone of each spin
states to the Moir\'{e} Brillouin zone. The inter-spin coupling coefficient
can be obtained as $\langle \phi _{\uparrow l\mathbf{k}_{\uparrow }}|\Omega
|\phi _{\downarrow l^{\prime }\mathbf{k}_{\downarrow }}\rangle =\delta _{%
\mathbf{q},\mathbf{q}^{\prime }}J_{\mathbf{g}_{\uparrow }\mathbf{g}%
_{\downarrow }}^{ll^{\prime }}(\mathbf{q})$, where $\mathbf{k}_{\uparrow }=%
\mathbf{q}+\mathbf{g}_{\uparrow }$ and $\mathbf{k}_{\downarrow }=\mathbf{%
q^{\prime }}+\mathbf{g}_{\downarrow }$, with $\mathbf{g}_{s}$ the Moir\'{e}
reciprocal lattice vectors and $\mathbf{q}$ the Moir\'{e} Bloch momentum
that is a good quantum number. We obtain the Hamiltonian Eq.~(1) in the main
text, which incorporates all range real-space tunnelings with high accuracy.
Another advantage of this momentum-space approach is that we only need to
keep $\mathbf{g}_{s}$ around the Dirac cones to correctly characterize the
low-energy physics at small twist angles, leading to rather rapid
convergence of the basis set.

Now we show how to evaluate the inter-spin coupling coefficients
\begin{eqnarray}
\langle \phi _{\uparrow l\mathbf{k}_{\uparrow }}|\Omega |\phi _{\downarrow
l^{\prime }\mathbf{k}_{\downarrow }}\rangle &=&\Omega \int d^{2}\mathbf{r}%
e^{-i\mathbf{k}_{\uparrow }\cdot \mathbf{r}}u_{\uparrow l\mathbf{k}%
_{\uparrow }}^{\ast }(\mathbf{r})e^{i\mathbf{k}_{\downarrow }\cdot \mathbf{r}%
}u_{\downarrow l^{\prime }\mathbf{k}_{\downarrow }}(\mathbf{r})  \nonumber \\
&=&\Omega \sum_{\mathbf{M}}e^{i(\mathbf{q^{\prime }}-\mathbf{q})\cdot
\mathbf{M}}\langle u_{\uparrow l\mathbf{k}_{\uparrow }}|e^{i(\mathbf{k}%
_{\downarrow }-\mathbf{k}_{\uparrow })\cdot (\mathbf{r-M})}|u_{\downarrow
l^{\prime }\mathbf{k}_{\downarrow }}\rangle _{\mathbf{M}}  \nonumber \\
&=&\Omega \delta _{\mathbf{q},\mathbf{q}^{\prime }}\langle u_{\uparrow l%
\mathbf{k}_{\uparrow }}|e^{i(\mathbf{g}_{\downarrow }-\mathbf{g}_{\uparrow
})\cdot \mathbf{r}}|u_{\downarrow l^{\prime }\mathbf{k}_{\downarrow
}}\rangle .
\end{eqnarray}%
Here $\mathbf{M}$ denotes the Moir\'{e} lattice vectors, and the term $%
\langle \cdots \rangle _{\mathbf{M}}$ in the second line (with integral over
the $\mathbf{M}$-th Moir\'{e} unit cell) is independent of $\mathbf{M}$. The
coefficients $J_{\mathbf{g}_{\uparrow }\mathbf{g}_{\downarrow }}^{ll^{\prime
}}(\mathbf{q})$ is
\begin{eqnarray}
J_{\mathbf{g}_{\uparrow }\mathbf{g}_{\downarrow }}^{ll^{\prime }}(\mathbf{q}%
) &=&\Omega \langle u_{\uparrow l\mathbf{k}_{\uparrow }}|e^{i(\mathbf{g}%
_{\downarrow }-\mathbf{g}_{\uparrow })\cdot \mathbf{r}}|u_{\downarrow
l^{\prime }\mathbf{k}_{\downarrow }}\rangle  \nonumber \\
&=&\Omega \sum_{\mathbf{p}_{\uparrow },\mathbf{p}_{\downarrow }}c_{\uparrow l%
\mathbf{k}_{\uparrow }}^{\mathbf{p}_{\uparrow }\ast }c_{\downarrow l^{\prime
}\mathbf{k}_{\downarrow }}^{\mathbf{p}_{\downarrow }}\int d^{2}\mathbf{r}%
e^{i(\mathbf{p}_{\downarrow }-\mathbf{p}_{\uparrow }+\mathbf{g}_{\downarrow
}-\mathbf{g}_{\uparrow })\cdot \mathbf{r}}  \nonumber \\
&=&\Omega \sum_{\mathbf{p}_{\uparrow },\mathbf{p}_{\downarrow },\mathbf{M}%
}c_{\uparrow l\mathbf{k}_{\uparrow }}^{\mathbf{p}_{\uparrow }\ast
}c_{\downarrow l^{\prime }\mathbf{k}_{\downarrow }}^{\mathbf{p}_{\downarrow
}}\frac{e^{i(\mathbf{p}_{\downarrow }-\mathbf{p}_{\uparrow }+\mathbf{g}%
_{\downarrow }-\mathbf{g}_{\uparrow })\cdot \mathbf{L}_{1}}-1}{(\mathbf{p}%
_{\downarrow }-\mathbf{p}_{\uparrow }+\mathbf{g}_{\downarrow }-\mathbf{g}%
_{\uparrow })\cdot \mathbf{L}_{1}}\frac{1-e^{i(\mathbf{p}_{\downarrow }-%
\mathbf{p}_{\uparrow }+\mathbf{g}_{\downarrow }-\mathbf{g}_{\uparrow })\cdot
\mathbf{L}_{2}}}{(\mathbf{p}_{\downarrow }-\mathbf{p}_{\uparrow }+\mathbf{g}%
_{\downarrow }-\mathbf{g}_{\uparrow })\cdot \mathbf{L}_{2}}  \label{eq:J} \\
&=&\Omega \sum_{\mathbf{p}_{\uparrow },\mathbf{p}_{\downarrow },\mathbf{M}%
}c_{\uparrow l\mathbf{k}_{\uparrow }}^{\mathbf{p}_{\uparrow }\ast
}c_{\downarrow l^{\prime }\mathbf{k}_{\downarrow }}^{\mathbf{p}_{\downarrow
}}\delta _{\mathbf{p}_{\downarrow }-\mathbf{p}_{\uparrow },\mathbf{g}%
_{\uparrow }-\mathbf{g}_{\downarrow }}.  \nonumber
\end{eqnarray}%
To obtain the last step, we have used $\mathbf{p}_{s}\cdot \mathbf{L}%
_{1}=2\pi (mp_{s1}+np_{s2})$ (similarly for $\mathbf{L}_{2}$), since $%
\mathbf{p}_{s}=p_{s1}\mathbf{e}_{s\mathbf{k}_{1}}+p_{s2}\mathbf{e}_{s\mathbf{%
k}_{2}}$ and $\mathbf{L}_{1}=m\mathbf{e}_{s1}+n\mathbf{e}_{s2}$, with $%
\mathbf{e}_{s1},\mathbf{e}_{s2}$ ($\mathbf{e}_{s\mathbf{k}_{1}},\mathbf{e}_{s%
\mathbf{k}_{2}}$) the primitive (reciprocal) lattice vectors of trap $V_{s}$.

\begin{figure}[b]
\includegraphics[width=1.0\linewidth]{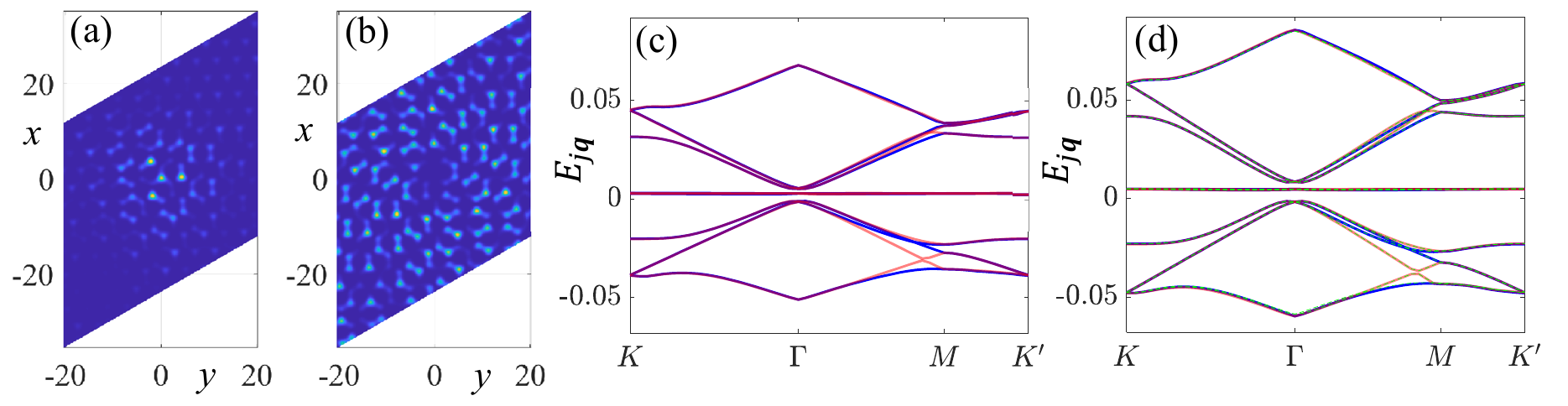}
\caption{(a) and (b) Typical density distributions of the Moire flat band
Bloch states $\protect\psi _{j\mathbf{q}}(\mathbf{r},\downarrow )$ in one
Moir\'{e} unit cell for $\mathbf{q}=K$ and $\mathbf{q}=\Gamma $,
respectively. Atoms are mainly distributed around the $AA$ region for $%
\mathbf{q}=K$ and become more uniformly distributed for $\mathbf{q}=\Gamma $%
. Other parameters are $\protect\theta =5.086^{\circ }$, $V_{0}=6E_{R}$ and $%
\Omega =0.116$. (c) Moir\'{e} bands at incommensurate twist angle $\protect%
\theta =4^{\circ }$ based on the generalized one-valley continuum model,
with $V_{0}=6E_{R}$ and $\Omega =0.0915$. The red and blue lines are bands
from $K_{s}$ and $K_{s}^{\prime }$ valleys, respectively. We have retained a
shell (containing 37 MBZs up to the third Moir\'{e} reciprocal lattice
vector shell) around the valley to construct the basis set. (d) Moir\'{e}
bands at commensurate twist $\protect\theta =5.086^{\circ }$ based on the
generalized one-valley continuum model, with $V_{0}=6E_{R}$ and $\Omega
=0.116$. The green thin dashed lines are Moir\'{e} bands obtained using the
full tight-binding basis.}
\label{fig:S2}
\end{figure}

With these coupling coefficients, we can diagonalize the single particle
Hamiltonian $H_{0}=H_{\uparrow }+H_{\downarrow }+H_{\uparrow \downarrow }$.
In the Moir\'{e} Bloch eigenbasis $\{\psi _{j\mathbf{q}}\}$, it reads
\begin{equation}
H_{0}=\sum_{j,\mathbf{q}}E_{j\mathbf{q}}\beta _{j\mathbf{q}}^{\dag }\beta _{j%
\mathbf{q}}.
\end{equation}%
The typical distributions of $\psi _{j\mathbf{q}}(\mathbf{r},s)$ are shown
in Figs.~\ref{fig:S2}a and \ref{fig:S2}b. Atoms are mainly distributed
around the $AA$ region for all $\mathbf{q}$ except a small area near $%
\mathbf{q}=\Gamma $, where atoms become more uniformly distributed. As a
result, the interaction and thereby the pairing is weak at $\Gamma $ point.

The above results and the Hamiltonian Eq.~(1) in the main text apply for any
commensurate twist angles. For incommensurate twist angles, there are no
well-defined Moir\'{e} patterns and Moir\'{e} bands if the twist angle is
too large. However, if the twist angle is small enough, Moir\'{e} patterns
can form even for incommensurate twist. In this case, the low-energy physics
is mainly determined by the states around the Dirac valleys, and the
inter-valley coupling is also negligible. Therefore, we can adopt a similar
approach as the continuum model by only keeping $\mathbf{g}_{s}$ around one
valley in the Hamiltonian, and the inter-spin couplings can be obtained
using Eq.~\ref{eq:J}. In Fig.~\ref{fig:S2}c, we plot the Moir\'{e} bands at
a small incommensurate twist angle using the generalized one-valley
continuum model mentioned above. We also plot the Moir\'{e} bands at a small
commensurate twist angle using the one-valley continuum model approach and
compare it with the results based on the full tight-binding basis (see Fig.~%
\ref{fig:S2}d). Their agreement confirms the validation of the one-valley
continuum model at small twist angles.

%To obtain the correct
%magic flatbands and low-energy physics, long range
%tunnelings beyond nearest neighbors should be taken into account

As we discussed in the main text, the energy scales of the magic physics is
small and at the order of $10^{-3}E_{R}$. Therefore, in the Wannier basis,
long-range (intra- and inter-spin) tunnelings beyond nearest neighbors that
are $\gtrsim 10^{-3}E_{R}$ should be taken into account to obtain correct
magic flat bands~\cite%
{PhysRevLett.99.256802S,PhysRevB.81.165105S,PNAS.108.12233S,PhysRevB.86.155449S}%
. In particular, the inter-spin site separations take various values and are
nearly continuously distributed for small twists, therefore the on-site, NN,
next-NN, etc., inter-spin tunnelings should be considered even when the
tunneling cutoff distance is small. The inter-spin tunnelings are also
%all tunnelings (especially the inter-spin couplings) are
highly anisotropic, which depend on both the relative orientation and
distance between the corresponding sites due to the three-fold rotational
symmetry of the Wannier orbitals. A small deviation in the tunnelling
coefficients may result in significant change in the band structures near
the `magic angle' due to the narrow bandwidths and approximate degeneracy of
the flat bands. The simplified real-space tight-binding model in~\cite%
{PhysRevA.100.053604S}, which assumes a simply isotropic Gaussian Wannier
function and includes only the nearest-neighbor intra-spin tunneling and
on-site inter-spin couplings, has significant deviation in determining the
bands at small twist angles.

To elaborate more on the effects of long-range real-space tunnelings, we
first solve for the maximally localized Wannier functions $w_{s,j}(\mathbf{r}%
)$ (with $j$ the site index) of the two spin-dependent lattices, then
calculate the band structures using the real-space tight-binding Hamiltonian
\[
H_{s}=\sum_{j,j^{\prime }}t_{jj^{\prime }}\alpha _{s,j}^{\dag }\alpha
_{s,j^{\prime }}
\]%
and
\[
H_{\uparrow \downarrow }=\sum_{j,j^{\prime }}J_{jj^{\prime }}\alpha
_{\uparrow ,j}^{\dag }\alpha _{\downarrow ,j^{\prime }}+h.c.,
\]%
where $\alpha _{s,j}^{\dag }$ is the creation operator of the Wannier
orbitals, and the tunneling coefficients are $t_{jj^{\prime }}=\int d^{2}%
\mathbf{r}w_{s,j}^{\ast }(\mathbf{r})[-\nabla _{\mathbf{r}}^{2}+V_{s}(%
\mathbf{r})]w_{s,j^{\prime }}(\mathbf{r})$ and $J_{jj^{\prime }}=\int d^{2}%
\mathbf{r}w_{\uparrow ,j}^{\ast }(\mathbf{r})\Omega w_{\downarrow ,j^{\prime
}}(\mathbf{r})$. For the lattice depth considered in this paper, we have $%
(t_{0};t_{1};t_{2};t_{3};t_{4};t_{5})=(-4.3458;-0.1967;0.0166;-0.0047;-0.0025;0.0009)E_{R}
$ for $V_{0}=6E_{R}$ and $%
(t_{0};t_{1};t_{2};t_{3};t_{4};t_{5})=(-2.2498;-0.2351;0.0245;-0.0089;-0.0045;0.002)E_{R}
$ for $V_{0}=4E_{R}$, with $t_{jj^{\prime }}=t_{\tilde{\imath}}$ if site $j$
is the $\tilde{\imath}$-th neighbor of site $j^{\prime }$ (where $t_{0}$ is
the on-site energy). Therefore, the cutoff distance for tunneling should be
up to the 5-th nearest neighbors of the bare hexagonal lattice. The
inter-spin coupling $J_{jj^{\prime }}$ is nearly continuously distributed
which decays exponentially with site separation (the on-site tunneling rate
is close to $\Omega $). In Figs.~\ref{figS:real_space_TB}a and ~\ref%
{figS:real_space_TB}b, we show the band structures for different cutoff
distance. We see that the flat band energy is largely shifted (by $\sim
0.25t_{1}$) for a cutoff only to the 1st nearest neighbors of the bare
hexagonal lattice. Nevertheless, magic flat bands are still observed with a
larger width (which can be reduced further by tuning $\Omega $), indicating
the robustness of the magic-angle physics.

\begin{figure}[b]
\includegraphics[width=0.95\linewidth]{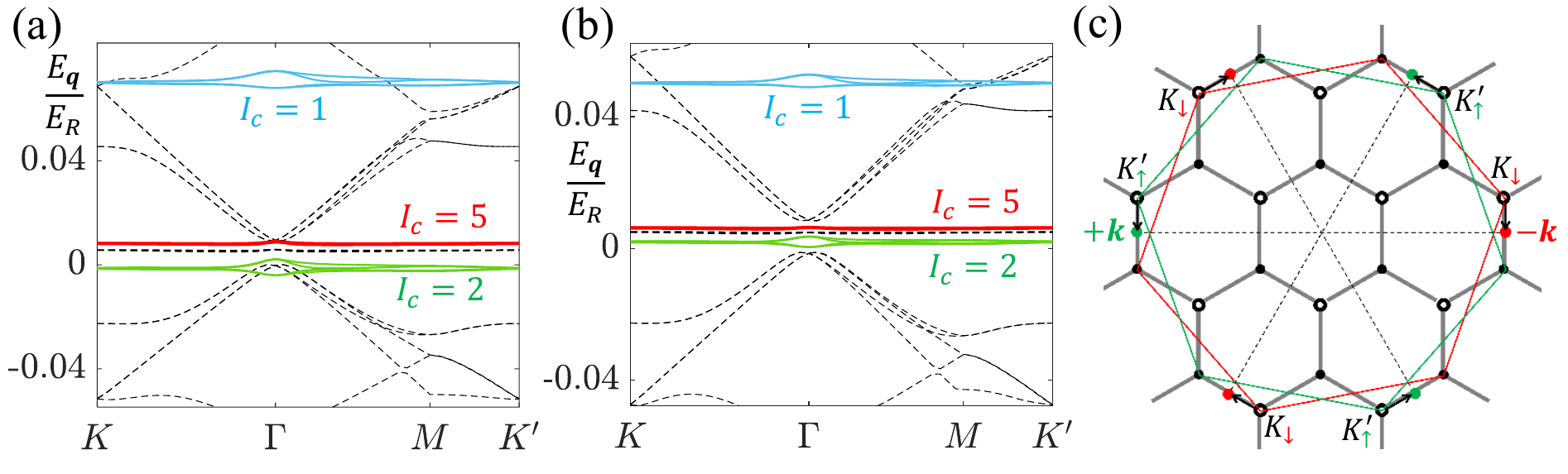}
\caption{(a) and (b) Moir\'{e} bands based on real-space tight-binding
Hamiltonian for $V_0=4E_R$ and $V_0=6E_R$, respectively. For cutoff distance
$I_c$, we keep the (intra- and inter-spin) tunnelings with site separation
not greater than the $I_c$-th neighbor separation of the bare hexagonal
lattice (e.g., we keep $t_{\tilde{\imath}}$ with $\tilde{\imath}\leq I_c$). The blue,
green and red solid lines correspond to the cutoff distance up to the $1$st,
$2$nd and $5$-th neighbors of the bare hexagonal lattice. The thin dashed
lines are the results of the momentum-space tight-binding Hamiltonian. Other
parameters are $\protect\theta =5.086^{\circ }$ and $\Omega =\Omega_f$. (c)
Schematic illustration of the pairing (indicated by dashed lines) in the
Brillouin zones, between $\uparrow$ states (green dots at $+\mathbf{k}$)
around valley $K^{\prime }_\uparrow$ and $\downarrow$ states (red dots at $-%
\mathbf{k}$) around valley $K_\downarrow$. The two pairing states are
located at the same side of the corresponding valleys (as indicated by the
black arrows), and thus the relative phase between $A$ and $B$ sublattice
sites (which is related to the chirality of the valley) leads to the
staggered pairing order.}
\label{figS:real_space_TB}
\end{figure}

\subsection{Superfluid orders}

The mean-filed interaction reads
\begin{eqnarray}
H_{\text{int}} &=&U_{0}\int d^{2}\mathbf{r}\hat{\Psi}_{\uparrow }^{\dag }(%
\mathbf{r})\hat{\Psi}_{\downarrow }^{\dag }(\mathbf{r})\hat{\Psi}%
_{\downarrow }(\mathbf{r})\hat{\Psi}_{\uparrow }(\mathbf{r})  \nonumber \\
&=&\int d^{2}\mathbf{r}[\hat{\Psi}_{\uparrow }^{\dag }(\mathbf{r})\hat{\Psi}%
_{\downarrow }^{\dag }(\mathbf{r})\Delta (\mathbf{r})+\hat{\Psi}_{\downarrow
}(\mathbf{r})\hat{\Psi}_{\uparrow }(\mathbf{r})\Delta ^{\ast }(\mathbf{r})-%
\frac{|\Delta (\mathbf{r})|^{2}}{U_{0}}]
\end{eqnarray}%
with local pairing amplitude $\Delta (\mathbf{r})=U_{0}\langle \hat{\Psi}%
_{\downarrow }(\mathbf{r})\hat{\Psi}_{\uparrow }(\mathbf{r})\rangle =\sum_{%
\mathbf{g}}\Delta _{\mathbf{g}}e^{i\mathbf{g}\cdot \mathbf{r}}$. We expand
the field operator in the Moir\'{e} Bloch basis $\hat{\Psi}_{s}(\mathbf{r}%
)=\sum_{j,\mathbf{q}}\beta _{j\mathbf{q}}\psi _{j\mathbf{q}}(\mathbf{r},s)$
and obtain the gap equation
\begin{equation}
\Delta _{\mathbf{g}}=U_{0}\sum_{j^{\prime },j,\mathbf{q}}\chi _{j^{\prime
}j}^{\mathbf{q}}(\mathbf{g})C_{\mathbf{q}}^{j^{\prime }j}[\Delta (\mathbf{r}%
)],  \label{eq:GapS}
\end{equation}%
where $\chi _{j^{\prime }j}^{\mathbf{q}}(\mathbf{g})=\frac{1}{L}\int d^{2}%
\mathbf{r}e^{-i\mathbf{g}\cdot \mathbf{r}}\psi _{j^{\prime }-\mathbf{q}}(%
\mathbf{r},\downarrow )\psi _{j\mathbf{q}}(\mathbf{r},\uparrow )$. The
correlation $C_{\mathbf{q}}^{j^{\prime }j}=\langle \beta _{j^{\prime }-%
\mathbf{q}}\beta _{j\mathbf{q}}\rangle $ can be obtained by solving the
Bogoliubov-de Gennes (BdG) Hamiltonian%
\begin{equation}
H_{\text{BdG}}=\sum_{j,\mathbf{q}}(E_{j\mathbf{q}}-\mu)\beta _{j\mathbf{q}}^{\dag
}\beta _{j\mathbf{q}}+\sum_{j,j^{\prime },\mathbf{q}}[\bar{\Delta}%
_{jj^{\prime }}(\mathbf{q})\beta _{j\mathbf{q}}^{\dag }\beta _{j^{\prime }-%
\mathbf{q}}^{\dag }+h.c.]
\end{equation}%
with $\bar{\Delta}_{jj^{\prime }}(\mathbf{q})=L\sum_{\mathbf{g}}\Delta _{%
\mathbf{g}}\chi _{j^{\prime }j}^{\mathbf{q}\ast }(\mathbf{g})$. We solve the
gap equation self-consistently by retaining only the four flat bands which
have much larger DOS than nearby bands.

As we discussed in the main text, at finite temperature, the relevant
physics is the Berezinskii-Kosterlitz-Thouless (BKT) transition since there
is no long-range superfluid order in two dimensions due to phase
fluctuations, and free vortex-antivortex pairs are formed spontaneously
below a critical temperature $T_{\text{BKT}}$~\cite{PhysRevLett.109.105302S,PhysRevLett.114.110401S}. The critical temperature $%
T_{c}$ by mean-field theory is often overestimated. Nevertheless, the
mean-field critical temperature $T_{c}$ provides an upper bound of $T_{\text{%
BKT}}$. Furthermore, the true BKT transition temperature could be obtained
from the mean-field superfluid weight $\rho _{s}$ which can be calculated
based on current response~\cite{PhysRevLett.112.086401S,julku2018superfluidS}%
. On the Moir\'{e} superlattice scale, our system corresponds to an $s$-wave
superfluid, we introduce a slowly varying superfluid phase $\Delta (\mathbf{r%
})\rightarrow \Delta (\mathbf{r})e^{i\Theta (\mathbf{r},\tau)}=\sum_{\mathbf{g}%
}\Delta _{\mathbf{g}}e^{i\mathbf{g}\cdot \mathbf{r}}e^{i\Theta (\mathbf{r}%
,\tau)}$, and the phase $\Theta (\mathbf{r},\tau)=\mathbf{Q}\cdot \mathbf{r}$
(with $\mathbf{Q}\rightarrow 0$) leads to a supercurrent $\mathbf{j}%
_{s}=\rho _{s}\mathbf{Q}$. The supercurrent can also be calculated using the
variation of the free energy $\mathbf{j}_{s}=\frac{1}{L}\langle \frac{\delta
H_{\text{BdG}}}{\delta \mathbf{Q}}\rangle $, the superfluid weight is
obtained by $\rho _{s}=\mathbf{j}_{s}/\mathbf{Q}$, and the BKT transition
temperature is given by $T_{\text{BKT}}=\frac{\pi }{2}\rho _{s}(T_{\text{BKT}%
})$~\cite{PhysRevLett.112.086401S,julku2018superfluidS}. Here the BdG
Hamiltonian reads
\begin{equation}
H_{\text{BdG}}(\mu ,\Delta ,\mathbf{Q})=\sum_{j,\mathbf{q}}(E_{j\mathbf{q}%
}-\mu)\beta _{j\mathbf{q}}^{\dag }\beta _{j\mathbf{q}}+\sum_{j,j^{\prime },%
\mathbf{q}}[\bar{\Delta}_{jj^{\prime }}(\mathbf{q,Q})\beta _{j\mathbf{q+Q/2}%
}^{\dag }\beta _{j^{\prime }\mathbf{-q+Q/2}}^{\dag }+h.c.]
\end{equation}%
where $\bar{\Delta}_{jj^{\prime }}(\mathbf{q,Q})=L\sum_{\mathbf{g}}\Delta _{%
\mathbf{g}}\chi _{j^{\prime }j}^{\mathbf{q,Q}\ast }(\mathbf{g})$ and $\chi
_{j^{\prime }j}^{\mathbf{q,Q}}(\mathbf{g})=\frac{1}{L}\int d^{2}\mathbf{r}%
e^{-i\mathbf{g}\cdot \mathbf{r}}\psi _{j^{\prime }\mathbf{-q+Q/2}}(\mathbf{r}%
,\downarrow )\psi _{j\mathbf{q+Q/2}}(\mathbf{r},\uparrow )$. The gap
equation is $\Delta _{\mathbf{g}}=U_{0}\sum_{j^{\prime },j,\mathbf{q}}\chi
_{j^{\prime }j}^{\mathbf{q,Q}}(\mathbf{g})C_{\mathbf{q,Q}}^{j^{\prime }j}$
with $C_{\mathbf{q,Q}}^{j^{\prime }j}=\langle \beta _{j^{\prime }\mathbf{%
-q+Q/2}}\beta _{j\mathbf{q+Q/2}}\rangle $.

Typically, $\rho _{s}(T)$ is
nearly a constant at low temperature $T$, and therefore we use the following
approximation $T_{\text{BKT}}\simeq \frac{\pi }{2}\rho _{s}(T=0)$. We
numerically calculate $T_{\text{BKT}}$ (the results are shown in Fig.~3a in
the main text) and find that $T_{\text{BKT}}\simeq 0.4T_{c}$. In particular,
we first set $\Theta (\mathbf{r},\tau)=0$ and solve for the order parameter $%
\Delta _{0}(\mathbf{r})$ with fixed chemical potential $\mu $, the
corresponding zero-temperature free energy is $\mathcal{F}(\mathbf{Q}%
=0)\equiv \frac{1}{L}\langle H_{\text{BdG}}(\mu ,\Delta _{0},\mathbf{Q}%
=0)\rangle $. Then we introduce the superfluid phase $\Theta (\mathbf{r},\tau)=%
\mathbf{Q}\cdot \mathbf{r}$ by setting the order parameter as $\Delta _{0}(%
\mathbf{r})e^{i\mathbf{Q}\cdot \mathbf{r}}$, and we calculate the free
energy $\mathcal{F}(\mathbf{Q})\equiv \frac{1}{L}\langle H_{\text{BdG}}(\mu
,\Delta _{0},\mathbf{Q})\rangle $. Since $\mathbf{j}_{s}=\rho _{s}\mathbf{Q}=%
\frac{\delta F(\mathbf{Q})}{\delta \mathbf{Q}}$ for $\mathbf{Q}\rightarrow 0$
at $T=0$, we have $F(\mathbf{Q})-F(0)=\frac{1}{2}\rho _{s}\mathbf{Q}^{2}=%
\frac{1}{2}\rho _{s}(\nabla \Theta )^{2}$ and the superfluid weight is
calculated as $\rho _{s}(T=0)=2\times \frac{F(\mathbf{Q})-F(0)}{\mathbf{Q}%
^{2}}$. The BKT critical temperature is obtained as $T_{\text{BKT}}\simeq
\pi \times \frac{F(\mathbf{Q})-F(0)}{\mathbf{Q}^{2}}$.

It can be verified that, the results based on current response are
consistent with those obtained through the effective action based on quantum
field theory. The partition function can be written as $Z=\text{Tr}%
(e^{-H/T})=\int \mathcal{D}[\bar{\Psi},\Psi ]e^{-S_{\text{eff}}[\bar{\Psi}%
,\Psi ]}$. The effective action is
\begin{equation}
S_{\text{eff}}[\bar{\Psi},\Psi ]=\int_{0}^{1/T}d\tau \left[ \int d^{2}%
\mathbf{r}\sum_{s}\bar{\Psi}_{s}\partial _{\tau }\Psi _{s}+\mathcal{H}(\bar{%
\Psi},\Psi )\right] ,
\end{equation}%
with $\mathcal{H}(\bar{\Psi},\Psi )=\int d^{2}\mathbf{r}\left\{ \sum_{s}[%
\bar{\Psi}_{s}(-\nabla ^{2}+V_{s}-\mu)\Psi _{s}+\bar{\Psi}_{s}\Omega \Psi _{\bar{%
s}}]+U_{0}\bar{\Psi}_{\uparrow }\bar{\Psi}_{\downarrow }\Psi _{\downarrow
}\Psi _{\uparrow }\right\} $ obtained by replacing field operator $\hat{\Psi}%
_{s}^{\dag }(\mathbf{r},t)$ and $\hat{\Psi}_{s}(\mathbf{r},t)$ with Grassman
field number $\bar{\Psi}_{s}$ and $\Psi _{s}$ in the Hamiltonian, and $\tau $
is the imaginary time. We can integrate out the quartic interaction term by
Hubbard-Stratonovich transformation, where the order parameter $\Delta $ is
defined. By further integrating out fermion fields, the partition function
reads $Z=\int \mathcal{D}[\Delta ^{\ast },\Delta ]e^{-S_{\text{eff}}[\Delta
^{\ast },\Delta ]}$ with
\begin{equation}
S_{\text{eff}}[\Delta ^{\ast },\Delta ]=\int_{0}^{1/T}d\tau \int d^{2}%
\mathbf{r}\frac{-|\Delta |^{2}}{U_{0}}-\frac{1}{2}\text{Tr ln}\left( i\omega
_{m}+M_{\text{BdG}}\right) +\frac{1}{T}\sum(\mathcal{E}_{sl\mathbf{k}_{s}}-\mu ).
\end{equation}%
Where $M_{\text{BdG}}$ is the BdG matrix in the Nambu basis, and $\omega _{m}
$ are Matsubara frequencies. For a static phase $\Theta (\mathbf{r},\tau)=%
\mathbf{Q}\cdot \mathbf{r}$, the effective action induced by the phase
fluctuation is
\begin{equation}
S_{\text{fluc}}=\frac{1}{2}\text{Tr ln}[i\omega _{m}+M_{\text{BdG}%
}(0)]-\frac{1}{2}\text{Tr ln}[i\omega _{m}+M_{\text{BdG}}(\mathbf{Q})].
\end{equation}
We can
further sum over Matsubara frequencies and obtain
\begin{equation}
S_{\text{fluc}}=\frac{1}{2%
}\sum_{j,\mathbf{q}}\text{ln cosh}\frac{E_{\text{BdG}}(j,\mathbf{q,0})}{2T}-\frac{%
1}{2}\sum_{j,\mathbf{q}}\text{ln cosh}\frac{E_{\text{BdG}}(j,\mathbf{q,Q})}{2T},
\end{equation}
with $E_{\text{BdG}}(j,\mathbf{q,Q})$ the BdG eigenfrequencies, which can be calculated numerically.
It is hard to proceed further to obtain an analytical
expression of $S_{\text{fluc}} (\mathbf{Q})$ due to the complicated setup.
Instead, we
assume  $S_{\text{fluc}}$ takes the form
\begin{equation}
S_{\text{fluc}%
}=\int_{0}^{1/T}d\tau \int d^{2}\mathbf{r}\frac{1}{2}\rho _{s}(\nabla\Theta)^{2}=\int_{0}^{1/T}d\tau \int d^{2}\mathbf{r}\frac{1}{2}\rho _{s}\mathbf{Q}^{2}
\end{equation}
for $\mathbf{Q}\rightarrow 0$ and a static phase $\Theta (\mathbf{r},\tau)=%
\mathbf{Q}\cdot \mathbf{r}$~\cite{PhysRevLett.109.105302S,PhysRevLett.114.110401S},
leading to $\rho _{s}=\frac{2T}{
\mathbf{Q}^{2}L}S_{\text{fluc}}$.
In the low temperature limit $T\rightarrow 0
$, we have
\begin{equation}\lim_{T\rightarrow 0}\frac{T}{L}S_{\text{fluc}}=\frac{1}{2L}\sum_{j,\mathbf{q}%
,E<0}[E_{\text{BdG}}(j,\mathbf{q,Q})-E_{\text{BdG}}(j,\mathbf{q,0})]=F(%
\mathbf{Q})-F(0).\end{equation}
Therefore, the results based on current response are
consistent with those obtained through the effective action. We have
numerically verified that, $\rho _{s}(T)=\frac{2T}{L\mathbf{Q}^{2}}S_{\text{fluc%
}}$ is nearly a constant at low temperature with $\rho _{s}(T_{\text{BKT}})\simeq\rho _{s}(T=0)$,
and $T_{\text{BKT}}=%
\frac{\pi }{2}\rho _{s}(T_{\text{BKT}})\simeq \frac{\pi }{2}\rho _{s}(T=0)$
is a very good approximation.

%We numerically calculate $T_{%
%\text{BKT}}$ (the results are shown in Fig.~3a in the main text) and find
%that $T_{\text{BKT}}\simeq 0.5T_{c}$. In particular, we first set $\Theta (%
%\mathbf{r},t)=0$ and solve for the order parameter $\Delta _{0}(\mathbf{r})$
%with fixed chemical potential $\mu $. The corresponding free energy is $%
%\mathcal{F}(0)=\frac{1}{L}\langle H_{\text{BdG}}(\mu ,\Delta _{0},\mathbf{Q}%
%=0)\rangle $. Then we introduce the superfluid phase $\Theta (\mathbf{r},t)=%
%\mathbf{Q}\cdot \mathbf{r}$ by setting the order parameter as $\Delta _{0}(%
%\mathbf{r})e^{i\mathbf{Q}\cdot \mathbf{r}}$, and calculate the free energy $%
%\mathcal{F}(\mathbf{Q})=\frac{1}{L}\langle H_{\text{BdG}}(\mu ,\Delta _{0},%
%\mathbf{Q})\rangle $. Since $\rho _{s}\mathbf{Q}=\frac{\delta F(\mathbf{Q})}{%
%\delta \mathbf{Q}}$ for $\mathbf{Q}\rightarrow 0$, we have $F(\mathbf{Q}%
%)-F(0)=\frac{1}{2}\rho _{s}\mathbf{Q}^{2}=\frac{1}{2}\rho _{s}(\nabla \Theta
%)^{2}$ and the superfluid weight is calculated as $\rho _{s}=2\times \frac{F(%
%\mathbf{Q})-F(0)}{\mathbf{Q}^{2}}$. Notice that $\rho _{s}$ dependent on the
%system temperature, therefore, the BKT critical temperature is obtained
%self-consistently through $T_{\text{BKT}}=\frac{\pi }{2}\rho _{s}(T_{\text{%
%BKT}})$.

The attractive $s$-wave interaction pairs atoms from opposite valleys of the
bare hexagonal lattice and the intra-sublattice pairing is dominant. Note
that the pairing is between Moir\'{e} momentum $\pm \mathbf{q}$, which is
mainly determined by the bare Bloch states $\phi _{\uparrow l\mathbf{k}}$
and $\phi _{\downarrow l-\mathbf{k}}$ nearest to the valleys (which
contribute most to the flat bands). Due to the relative twist, $\pm \mathbf{k%
}$ are located at the same side of the corresponding valleys (as illustrated
in Fig.~\ref{figS:real_space_TB}c). Therefore, we have $\phi _{\uparrow l%
\mathbf{k}}\propto \lbrack 1,e^{i\gamma _{\uparrow \mathbf{k}}}]^{T}$ and $%
\phi _{\downarrow l-\mathbf{k}}\propto \lbrack 1,e^{i\gamma _{\downarrow -%
\mathbf{k}}}]^{T}$ in the $A$ and $B$ sublattice basis, with the relative
phases $\gamma _{\uparrow \mathbf{k}}\simeq -\gamma _{\downarrow -\mathbf{k}%
}+\pi $ (which are related to the chirality of the valleys, i.e., the Berry
phase on loops surrounding the valley). The relative phase $\gamma _{s%
\mathbf{k}}$ between $A$ and $B$ sublattice sites are responsible for the
Larkin-Ovchinnikov phase with staggered pairing order $\Delta (\mathbf{r}%
)\propto \langle \phi _{\uparrow l\mathbf{k}}\phi _{\downarrow l-\mathbf{k}%
}\rangle \propto \lbrack 1,-1]^{T}$.

\begin{figure}[tb]
\includegraphics[width=1.0\linewidth]{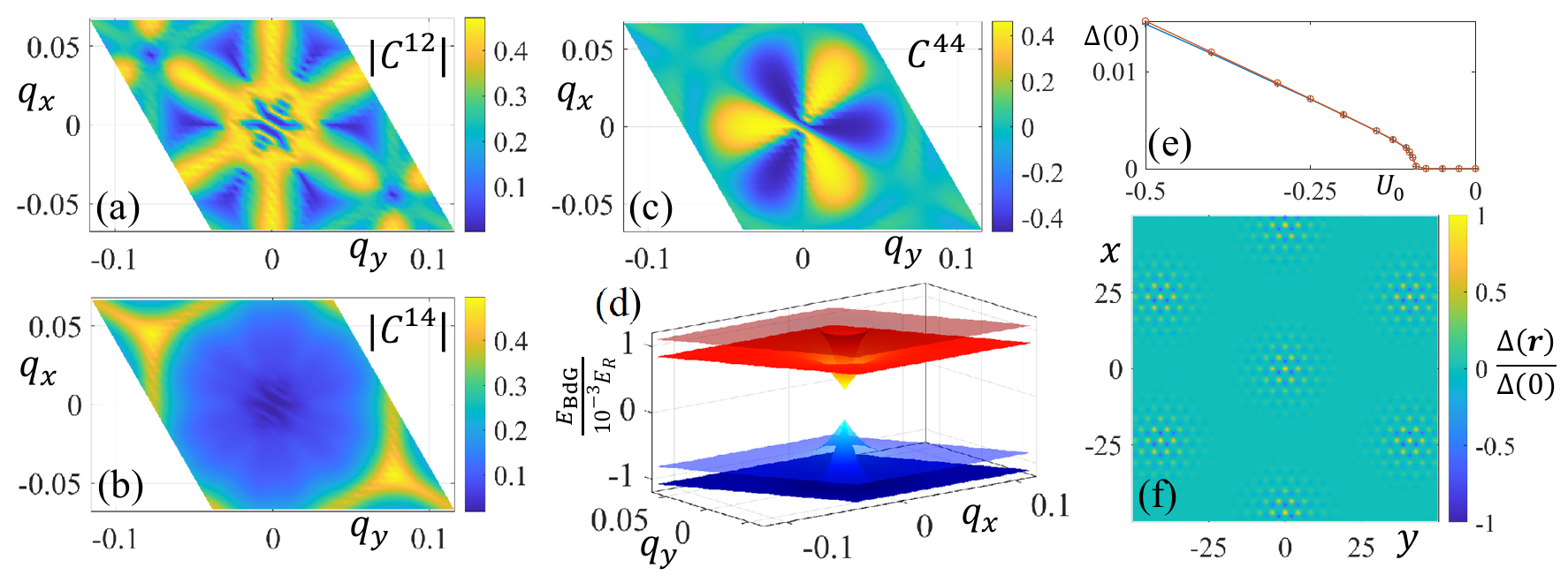}
\caption{(a)-(c) The correlations $C_{\mathbf{q}}^{j^{\prime }j}$ for
different $j^{\prime },j$ with $U_{0}=-0.5$. (d) The superfluid band
structures (two middle particle/hole bands are not shown). (e) Superfluid
order $\Delta (\mathbf{r}=0)$ as a function of $U_{0}$ obtained by retaining
12 bands (circles) and 4 bands (plus signs) with $\protect\mu =0.00485$. (f)
Superfluid order $\Delta (\mathbf{r})$ at $U_{0}=-0.5$. Common parameters: $%
\protect\theta =5.086^{\circ }$, $V_{0}=6E_{R}$ and $\Omega =0.116$.}
\label{fig:S3}
\end{figure}

\begin{figure}[b]
\includegraphics[width=1.0\linewidth]{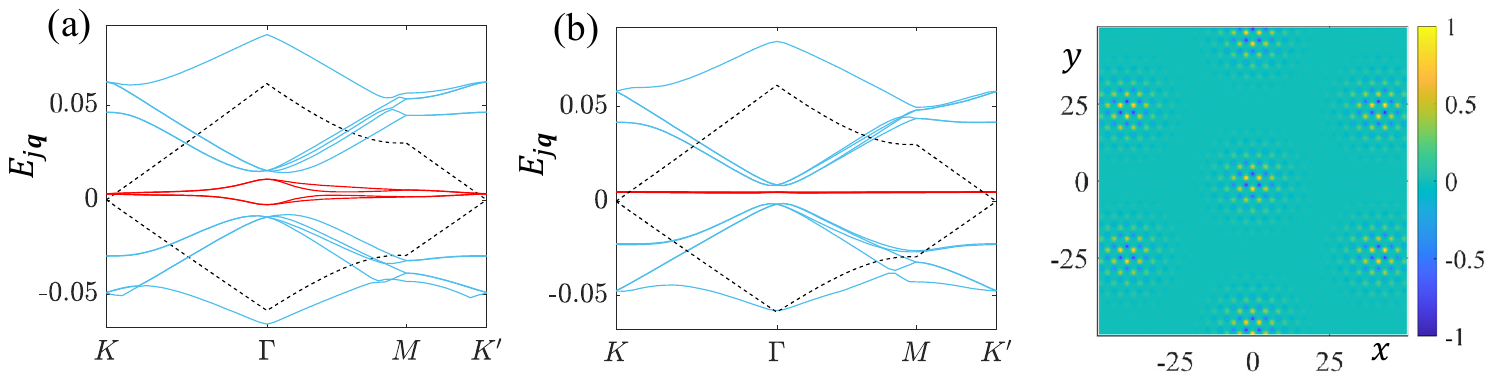}
\caption{(a) and (b) Moir\'{e} bands for $AB$ stacking with twist axis at
one coinciding site (where $A$ site of $V_{\uparrow }$ coincides with $B$
site of $V_{\downarrow }$). All other parameters in (a) and (b) are the same
as Figs.~2a and 2b in the main text, respectively. The bands are almost
identical with that in Figs.~2a and 2b in the main text, except that the two
valence (conduction) flat bands become degenerate at $K$ and $K^{\prime }$
points and a tiny gap develops between the valence and conduction flat
bands. (c) The normalized pairing order $\Delta (\mathbf{r})$ (maximum is
normalized to 1) starting from $AA$ stacking with twist axis at one hexagon
center. Other parameters are $\protect\theta =5.086^{\circ }$, $V_{0}=6E_{R}$%
, $\Omega =\Omega _{f}=0.116$ and $U_{0}=-0.5$.}
\label{fig:S4}
\end{figure}

The correlations $C_{\mathbf{q}}^{j^{\prime }j}=\langle \beta _{j^{\prime }-%
\mathbf{q}}\beta _{j\mathbf{q}}\rangle $ for different $j,j^{\prime }$ are
shown in Figs.~\ref{fig:S3}a-c. The pairing is mainly between bands from
different valleys, and the conduction bands from different valleys become
degenerate along the high symmetric $\Gamma $-$K$ lines with avoided
crossing (a tiny gap) due to inter-valley couplings. As we discussed in the
main text, $C_{\mathbf{q}}^{11}$ changes from characterizing $K_{s}$-$K_{%
\bar{s}}^{\prime }$ to $K_{s}^{\prime }$-$K_{\bar{s}}$ correlations;
therefore, $C_{\mathbf{q}}^{11}$ changes the sign across the $\Gamma $-$K$
lines where $C_{\mathbf{q}}^{12}$ is mainly distributed.
% around the $\Gamma$-$K$ lines.
This means that though $C_{\mathbf{q}}^{j^{\prime }j}$ show $f$-wave
structures, their combined effects lead to the nearly uniform superfluid gap
(see Fig.~\ref{fig:S3}d) and the pairing is $s$-wave. The small superfluid
gap (weak pairing) at $\Gamma$ is due to the nearly uniform Moir\'{e}
wavefunction there (see Fig.~\ref{fig:S2}b). Moreover, we also calculate the
results by including 8 nearby Moir\'{e} bands (12 bands in total with 4 flat
bands, 4 higher and 4 lower bands) for comparison, and find that these
nearby bands have very little effects on the phase diagram and superfluid
order (as shown in Figs.~\ref{fig:S3}e and \ref{fig:S3}f).

In the above mean-field approach, we have assumed a real-space pairing
order. Alternatively, we can assume the momentum-space pairing order. In
particular, we first write the interaction Hamiltonian as
\[
H_{\text{int}}=\sum U_{j_{1}j_{2};j_{3}j_{4}}^{\mathbf{Q;q,q^{\prime }}%
}\beta _{j_{1}\mathbf{Q+q}}^{\dag }\beta _{j_{2}\mathbf{Q-q}}^{\dag }\beta
_{j_{3}\mathbf{Q-q^{\prime }}}\beta _{j_{4}\mathbf{Q+q^{\prime }}}
\]%
with $U_{j_{1}j_{2};j_{3}j_{4}}^{\mathbf{Q;q,q^{\prime }}}=U_{0}\int d^{2}%
\mathbf{r}\psi _{j_{1}\mathbf{Q+q}}^{\ast }(\mathbf{r},\uparrow )\psi _{j_{2}%
\mathbf{Q-q}}^{\ast }(\mathbf{r},\downarrow )\psi _{j_{3}\mathbf{Q-q^{\prime
}}}(\mathbf{r},\downarrow )\psi _{j_{4}\mathbf{Q+q^{\prime }}}(\mathbf{r}%
,\uparrow )$. Here $\mathbf{Q}$, $\mathbf{q}$ and $\mathbf{q^{\prime }}$ are
superlattice momenta in the Moir\'{e} BZ. We restrict the interaction to the
$\mathbf{Q}=0$ BCS channel, and assume the momentum-space order $\bar{\Delta}%
_{j_{1}j_{2}}(\mathbf{q})=\sum_{j_{3},j_{4},\mathbf{q}^{\prime
}}U_{j_{1}j_{2};j_{3}j_{4}}^{\mathbf{0;q,q^{\prime }}}C_{\mathbf{q}^{\prime
}}^{j_{3}j_{4}}$. The correlation $C_{\mathbf{q}^{\prime }}^{j_{3}j_{4}}$
can be obtained by solving the BdG Hamiltonian%
\begin{equation}
H_{\text{BdG}}=\sum_{j,\mathbf{q}}(E_{j\mathbf{q}}-\mu)\beta _{j\mathbf{q}}^{\dag
}\beta _{j\mathbf{q}}+\sum_{j_{1},j_{2},\mathbf{q}}[\bar{\Delta}%
_{j_{1}j_{2}}(\mathbf{q})\beta _{j_{1}\mathbf{q}}^{\dag }\beta _{j_{2}-%
\mathbf{q}}^{\dag }+h.c.],
\end{equation}%
which allows us to obtain the superfluid order $\bar{\Delta}_{j_{1}j_{2}}(%
\mathbf{q})$ self-consistently. Using this approach, we calculate superfluid
order by keeping only the four flat bands and find that the orders $\bar{%
\Delta}_{j_{1}j_{2}}(\mathbf{q})$ are the same (up to tiny numeric errors)
as those obtained by assuming a real-space pairing order [i.e., $%
\sum_{j_{3},j_{4},\mathbf{q}^{\prime }}U_{j_{1}j_{2};j_{3}j_{4}}^{\mathbf{%
0;q,q^{\prime }}}C_{\mathbf{q}^{\prime }}^{j_{3}j_{4}}\simeq L\sum_{\mathbf{g%
}}\Delta _{\mathbf{g}}\chi _{j_{1}j_{2}}^{\mathbf{q}\ast }(\mathbf{g})$].
The two approaches lead to the same superfluid phase, correlation $C_{%
\mathbf{q}}^{j_{3}j_{4}}$, as well as $\Delta (\mathbf{r})$ (which is
determined by $C_{\mathbf{q}}^{j_{3}j_{4}}$).
%The real-space pairing order can be obtained from $%
%C^{j_3j_4}_{\mathbf{q}}$,
%as shown in Fig.~S3.

\subsection{Effects of different stackings and twist axes}

We have focused on the twists starting from $AA$ stacking with the twist
axis at one sublattice site. Like the magic behaviors in TBG~\cite%
{PhysRevLett.99.256802S,PhysRevB.81.165105S,PNAS.108.12233S,PhysRevB.86.155449S}%
, here the twist axis or stacking position do not affect the appearance of
magic flat bands at small twist angles, as shown in Figs.~\ref{fig:S4}a and %
\ref{fig:S4}b. When $\theta $ is small, the Moir\'{e} bands for different
stackings and twist axes are almost identical. For $AB$ stacking with twist
axis at one coinciding site (where $A$ site of $V_{\uparrow }$ coincides
with $B$ site of $V_{\downarrow }$), the two valence (conduction) flat bands
become degenerate at $K$ and $K^{\prime }$ points and a tiny gap develops
between the valence and conduction flat bands. We find that the Moir\'{e}
bands for $AB$ ($AA$) stacking with twist axis at one hexagon center is the
same as that for $AA$ ($AB$) stacking with twist axis at one coinciding
site. For different stackings and twist axes, the superfluid orders and the
phase diagrams are similar, where the pairing order is staggered and
distributed mainly around the $AA$ region (see Fig.~\ref{fig:S4}c).

\subsection{Difference with twisted bilayer graphene}

%Though our spin-twisted optical lattices have many similarities with the TBG
%system, there are several important differences worth reemphasizing.
In this section, we reemphasize some of the differences between our system
and the TBG system.

(1) The two twisted lattice potentials are state dependent, and one
potential does not affect atoms trapped by the other. This is different from
the electrons in TBG, where electrons in one layer can feel the potential of
atoms in the other layer.

(2) Our system is physically a single-layer system and we twist the lattice
for atomic (pseudo-)spin states (i.e., atomic internal energy levels). The $%
z $-direction is tightly confined by an additional state-independent
potential using the so-called magic-wavelength lasers. Therefore, the two
spin states have identical Wannier orbital along the $z$ direction. The
inter-spin tunnelings, realized by additional lasers, are different from the
inter-layer tunnelings between $p_{z}$ orbitals~\cite%
{PhysRevLett.99.256802S,PhysRevB.81.165105S} in TBG where a large
inter-layer distance exist. The existence of magic behaviors in our system
is not a straightforward derivative of TBG.

(3) The optical lattice potential here takes a cosine form which is much
simpler comparing to the atomic potential in graphene. Therefore the bare
bands and inter-spin couplings can be obtained accurately by directly
solving for the Bloch states $\phi _{sl\mathbf{k}_{s}}$ in our system. While
for the TBG, real-space tight-binding approximation based on Slater-Koster
parameters is usually adopted~\cite%
{PhysRevLett.99.256802S,PhysRevB.81.165105S}.

(4) Long-range tunnelings are more significant in our system because the
optical lattices considered here are relatively shallow. A shallow lattice
not only improves the atomic lifetime (through reducing the atomic decay
rate), but also increases the bare Dirac velocity $v_{D}$ (a larger $v_{D}$
leads to larger gaps and better isolation of the flat bands).

(5) The interactions are dominated by the $s$-wave scattering between atoms
in different spin states that are coupled and relatively twisted. In TBG,
the electronic interactions are more complex and include both Coulomb
repulsive interaction and/or phonon-mediated attractive interactions, which
mainly involve electrons in the same layer with no relative twist~\cite%
{PhysRevB.98.220504S,PhysRevLett.122.257002S,PhysRevLett.121.257001S,
PhysRevX.8.041041S,PhysRevLett.121.217001S,PhysRevLett.122.026801S}. The
unique interaction in our system can lead to interesting Larkin-Ovchinnikov
superfluid orders that do not exist in TBG.

(6) Finally, the advantage of cold atom system is that the parameters (e.g.,
inter-spin tunnelings, lattice depth, lattice constant, interactions, etc.)
are highly tunable. This not only leads to magic behaviors in a wide range
of parameter space, but also opens various possibilities for exploring novel
twistronics in cold atom systems.

\end{widetext}

\end{document}